\documentclass{article}

\usepackage{PRIMEarxiv}

\usepackage[utf8]{inputenc} 
\usepackage[T1]{fontenc}    
\usepackage{hyperref}       
\usepackage{url}            
\usepackage{booktabs}       
\usepackage{amsfonts}       
\usepackage{nicefrac}       
\usepackage{microtype}      
\usepackage{lipsum}
\usepackage{fancyhdr}       
\usepackage{graphicx}       
\graphicspath{{media/}}     
\usepackage{amsmath}
\usepackage{cite}
\usepackage{amsmath,amssymb,amsfonts}
\usepackage{algorithmic}
\usepackage{graphicx}
\usepackage{textcomp}
\usepackage{xcolor}
\usepackage{url}
\usepackage{hyperref}
\usepackage{booktabs}
\usepackage{authblk}
\usepackage{multicol}
\title{Chiplets on Wheels : Review Paper on holistic chiplet solutions for autonomous vehicles
}

\author[1]{Swathi Narashiman\thanks{Emails: \texttt{rudhranarashiman.225@gmail.com}, \texttt{2004venk@gmail.com}, \texttt{1724akul@gmail.com}, \texttt{deepaksridhar13@gmail.com}, \texttt{harishrajesh2002@gmail.com}, \texttt{sanjaysattvaxyz@gmail.com}, \texttt{sriram.aniruddha@gmail.com}, \texttt{ee20b050@smail.iitm.ac.in}, \texttt{ee20b149@smail.iitm.ac.in}, \texttt{ragucvachess@gmail.com}}}
\author[1]{Venkat A}
\author[2]{Divyaratna Joshi}
\author[3]{Deepak Sridhar}
\author[1]{Harish Rajesh}
\author[1]{Sanjay Sattva}
\author[1]{Aniruddha S}
\author[1]{Jayanth B}
\author[1]{Varun Manjunath}
\author[4]{Ragavendiran N}

\affil[1]{Department of Electrical Engineering, Indian Institute of Technology, Madras}
\affil[2]{Department of Chemical Engineering,Indian Institute of Technology, Madras }
\affil[3]{Department of Metallurgical and Materials Engineering, Indian Institute of Technology, Madras}
\affil[4]{Department of Mechanical Engineering, Indian Institute of Technology, Madras}

\begin{document}
\maketitle

\begin{abstract}
On the advent of the slow death of Moore's law, the silicon industry is moving towards a new era of chiplets.The automotive industry is experiencing a profound transformation towards software-defined vehicles, fueled by the surging demand for automotive compute chips, expected to reach $20-22$ billion by 2030. High-performance compute (HPC) chips become instrumental in meeting the soaring demand for computational power. Various strategies, including centralized electrical and electronic architecture and the innovative Chiplet Systems, are under exploration. The latter, breaking down System-on-Chips (SoCs) into functional units, offers unparalleled customization and integration possibilities. The research accentuates the crucial open Chiplet ecosystem, fostering collaboration and enhancing supply chain resilience. In this paper, We address the unique challenges that arise when we attempt to leverage chiplet-based architecture to design a holistic silicon solution for the automotive industry. We propose a throughput-oriented micro-architecture for ADAS \& infotainment system alongside a novel methodology to evaluate chiplet architectures. Further, We develop in-house simulation tools leveraging the \textbf{gem5} framework to simulate for latency and throughput.Finally, We perform an extensive design of thermally-aware chiplet placement and develop a micro-fluids based cooling design. 
\footnote{
\href{https://github.com/akulsylvania/Chiplet-on-Wheels/tree/main}{GitHub Link to simulation files } }
\end{abstract}


\section{Introduction}
The automotive industry is undergoing a significant transformation as it shifts towards software-defined cars, driven by the growing demand for automotive compute chips. The automotive compute market is expected to grow to \$20 billion to \$22 billion in 2030 \cite{b1}. This shift is leading to a change in focus from hardware to software within the industry. To meet the increasing demand for compute power, high-performance compute (HPC) chips are crucial, as automakers seek a resilient supply chain for customizable System-on-Chips (SoCs), while chip suppliers aim to recoup their upfront research and development (R\&D) investments.

Multiple approaches are being considered to transition towards a single compute stack, including a gradual migration to a centralized electrical and electronic (E/E) architecture, employing domain controllers, customized SoCs, and Chiplet Systems. Chiplet Systems involve breaking down SoCs into functional Chiplets, offering greater customization and integration possibilities, ultimately supporting future centralized compute stacks.

An open ecosystem of Chiplet Systems can create a new value chain dynamic within the automotive compute market, which is projected to grow significantly. This open ecosystem enables various stakeholders, including automakers, Tier-1 suppliers, and smaller chip suppliers, to collaborate in chip development, enhancing supply chain resilience and reducing the risk of lock-in effects. Overall, this shift in the automotive compute industry holds the potential for various players to benefit from these new dynamics and opportunities.

In this context, We aim to perform a holistic analysis of adoption of Chiplet technology in the automotive industry. The paper is organized as follows: Section \ref{A} gives an overview of the automotive E\&E landscape, evolving compute demands and the challenges ahead. Section \ref{B} first describes the qualitative benefits of chiplets over a SoC and then proceeds to quantitatively evaluate the differences between SoCs and Chiplets in terms of cost, power and Performance. Next We propose 2 areas of application of Chiplets in the automotive compute stack, giving a overview of special purpose computing. Then we give a detailed analysis of leading communication technologies and the interconnect IPs that are used in the industry. Finally We propose a thermal and packaging solution for the chiplet architecture we propose.

We will also explore the cutting-edge technologies in the field of chiplet design, specifically focusing on novel interconnect technologies and communication protocols. Our investigation will compare these advancements to traditional monolithic System-on-Chip (SoC) architectures. To provide a comprehensive analysis, we will delve deeper into various interconnect protocols, examining their latency and communication efficiency. This examination will be supported by simulations as well as insights derived from a synthesis of research papers and technical articles.

In final sections of the paper, We present a thermal design methodology, using simulation tools that optimize the complex dependencies involved in packaging a chip. We design a cooling solution based on microfluidics, giving special considertions to the unique constraints posed by the automotive environment.

\section{Automotive E\&E Architecture}\label{A}
\subsection{Overview}
E/E(electricals/electronics) architecture defines the arrangement of E/E components (e.g., Sensors, Controllers, or Actuators) connected by Communication components (e.g., buses), orchestrated and supervised by compute resources.\cite{b19} 
\subsection{Evolving Compute Demands} 
The automotive industry has been evolving its E/E architectures to manage the complexity of future vehicle systems. The current trend is to move away from domain-specific decentralized models towards a cross-domain, centralized E/E architecture that uses only a few powerful vehicle computers instead of many individual control units. This approach is called a vehicle-centralized, zone-oriented E/E architecture, which uses zone ECUs to connect the vehicle computers to the remaining embedded control units, sensors, and actuators. This architecture reduces the number of control units, enabling more complex cross-domain functions, leading to lower system complexity and increased security.\cite{b20}

\subsection{Challenges}
The automotive industry finds itself in the midst of a transformative era, with rapid technological advancements reshaping the way we perceive the automotive world. This dynamic environment brings both opportunities and challenges to automotive E/E architecture. Future E/E systems must exhibit flexibility to rapidly introduce new innovations and facilitate software sharing. To achieve this, open standards, modular components, and well-defined interfaces are vital. External communication, while enhancing the driving experience, leads to higher data traffic and security risks, necessitating the implementation of robust cybersecurity measures and secure communication protocols. The challenge of computing power arises as serial computing in embedded systems nears its limits, demanding the incorporation of high-performance computing units and cloud computing integration. Communication bandwidth is another issue, with inter-domain and cross-domain communication requiring improvements to support future data traffic. Lastly, scalability is a multifaceted challenge, influenced by diverse market segments and technologies. It calls for scalable architectures to handle complex and expensive variant management, ensuring adaptability to varying market demands. Addressing these challenges will be pivotal in shaping the future of automotive E/E architecture.\cite{b20}

\subsection{Overview of the Architecture}

\begin{figure}[htbp]
  \centering
  \includegraphics[width=0.8\textwidth]{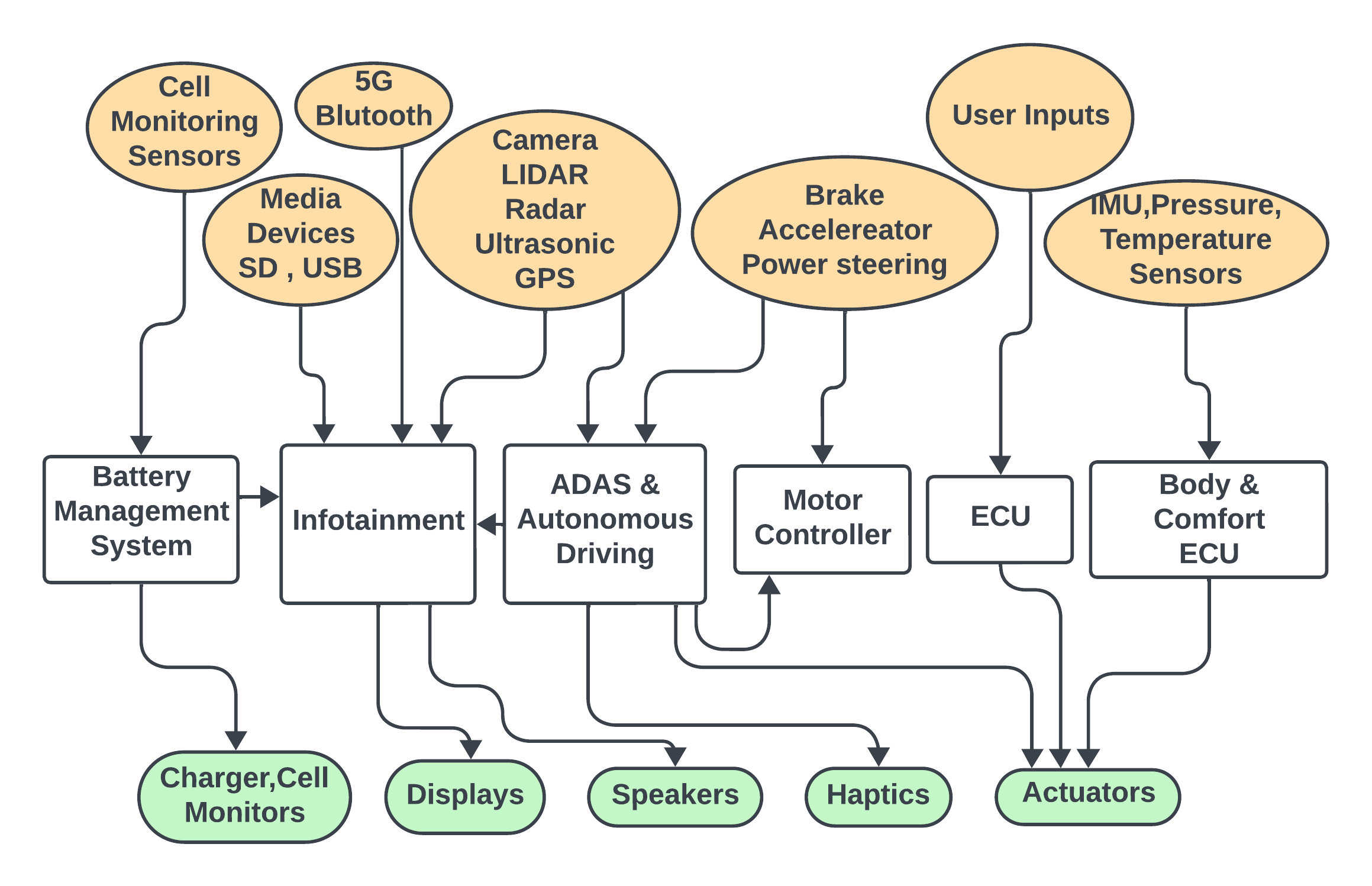}
  \caption{A High-Level E\&E Architecture : The chart above provides a high-level schematic representation of the input, output, and task categories within E\&E architectures}
\end{figure}
Apart from ADAS \& AD and Infotainment Block, other blocks require a compute performance similar to a general purpose microcontroller and from the inputs the sensors required for the AD processing are a significantly more demanding input than others.
\subsubsection{Points of interest}
\paragraph{I/O Received and Processed by ADAS and AD}
Advanced Driver Assistance Systems (ADAS) and Autonomous Driving (AD) systems rely on diverse sensors, including high-resolution cameras, LIDAR units, radar sensors, ultrasonic sensors, and GPS data, each contributing unique information to facilitate precise decision-making. Depending on the sensor, the I/O throughput requirements can vary from 10 Mbps to 10 Gbps, contingent upon factors such as resolution, frame rate, and compression ratios Therefore, when we combine multiple sensors, the total throughput can range from tens to hundreds of Gbps.In this context, low latency is imperative, with a general target of less than 100 ms for most scenarios and a preference for less than 50 ms in high-speed or emergency situations, ensuring rapid and accurate response to the dynamic environment.\cite{b18} \cite{b21}

\paragraph{The Heavy Computing Requirements of ADAS and AD}
A Level 4 AD system may need 16 to 32 GB of RAM, 64 to 128 GB of non-volatile memory, a CPU with 16 to 32 cores operating at 2 to 4 GHz, and 200 teraflops of computing power. To put this into perspective, this level of computing power is roughly equivalent to the combined computing capacity of 100 high-end laptops.\cite{b22} \cite{b23}

\paragraph{Infotainment I/O and Compute Needs}
This involves managing diverse I/O sources, including sensors, Wi-Fi, Bluetooth, media devices, 5G , ADAS \& AD and other Blocks. It also requires computing capabilities supporting features like personalization, AI-based assistants, voice recognition, over-the-air updates, video streaming, and gaming, comparable to a desktop.

\section{Analysis of Chiplets}\label{B}
Under this section, we analyse the key differences between Chiplets and System on Chips(SoCs) and mathematically analyze the scenarios where Chiplets are clear winners when we optimize 3 important factors of a semiconductor chip - cost, performance and power.

\subsection{Factors of Optimization}
An SoC can be split up into following functional parts: Logic, Memory, Analog Components. The overall performance of an SoC can be improved  by either increasing the clock speed and thereby increasing the FLOPS or increasing the size of compute and corresponding memory (Cache) and so that the micro-controller can access and process data quickly. The first approach is not preferred because the crystal clock speeds(XTALs) cannot be increased beyond a specified limit (saturated to about 6GHz). And the bottleneck with the latter is the Moore's law and the very fact that an entire SoC is being fabricated on a single process node results in compromise of quality as both logic and memory cannot be manufactured to their best on the same process node. On this note \textbf{Divide and Conquer Approach} is considered to be the state of art in manufacturing ICs. When this is done, Analog components can be fabricated on the cheapest and largest process node, Logic can be fabricated on the latest process node and the memory on the smallest die. The smaller the size of the die, defects are less likely to occur and the more is the yield.

\subsection{Generic Chiplet Consolidation}
Chiplet Consolidation is an indispensable step in identifying the best decomposition of a complex SoC. In general, an SoC can be split into 4 Chiplets integrated together by appropriate communication modes:
\begin{enumerate}
    \item The Logic Corelet: This contains the cores of the chiplet and L1 Cache Memory.
    \item The Input-Output Dielet: This consists if the Input-Output Modules as a Chiplet. The I/O Memory Space can also be incorporated into this Chiplet.
    \item The Memory Dielet: This comprises of L2, L3 cache and the Peripheral Component Interconnects(PCI).
    \item The GPU Dielet: This contains the integrated GPU.
\end{enumerate}

\subsection{Facets of a Chiplet}
\begin{itemize}
    \item Each Dielet can be individually fabricated on a suitable process node.
    \item When SoCs have to completely replaced for upgrading/debugging, a malfunctioning  Chiplet can be isolated and replaced while keeping the retaining the rest of the architecture. This also means that a Chiplet package is partially functional and can be utilised to some degree.
    \item  A heterogeneous fabricating architecture can be employed where some Chiplets are designed in-house and others are procured.
    \item Chiplets are found to be extremely effective for computationally intensive AI models deployed in autonomous cars.  Instead of slow and power hungry DRAM access, it is better to keep a dedicated memory as close as possible to the compute core. In our proposed Chiplet consolidation, we recommend that we place the L1 Cache in the Logic Corelet in order to speed up the computation.
    \item Chiplets resolves supply chain bottlenecks for OESMs that manufactures hardware architecture with millions of components embedded in Die sized chips. A heterogeneous planning strategy can help in parallel production and refinement of chiplets.
\end{itemize}

\subsection{Ideal Size of a Chiplet}
While we argue that decomposing SoCs to Chiplets is a better solution, research shows that there is a tight upper cap on the size of Chiplet. This is because when an SoC is dissolved into Chiplets of tinier sizes, the number of Chiplets increases and so is the  requirement of  I/O cells and this  subsequently demands potential Electrostatic Discharge(ESD) protection. Hence, cost cannot be optimal for Chiplets of smaller sizes ,also this affects the amount of memory a Chiplet posses and increases the latency in transmit of data through the interconnects. 

The optimal die size  appears to be 50-150mm2 in 40nm process node and 40-80mm2 in 7nm process node for microprocessor type logic. For random logic, the optimal size is close to 200mm2 \cite{10247947}.
\subsection{Modeling Cost,Power and Performance}
\subsubsection{{{Cost}}}
Cost is modelled as follows
\begin{equation}
cost = costPerDie/ assemblyYield
\end{equation}
\begin{equation}
\text{assemblyYield} =0.999^{Ndie} * 0.999999 ^ {Ng}
\end{equation}
Where Ndie is the number of dies made out of a 300mm wafer and Ng is the number of gates . We assume that the size on an average is 40mm2 for the computation of Ng
\begin{equation}
\text{costPerDie} = \text{costPerWafer}/\text{dieYield}
\end{equation}
\begin{equation}
dieyield=(1+d_0 x/\alpha)^{-\alpha}
\end{equation}
where, $d_0$ is a constant and $\alpha$ is the cluster parameter, x is the area of the die.
For an SoC we assume that the area is 858mm2 and for a chiplet we assume that the split is in such a way that the resultant chiplet has a size of 170mm2 approximately.
Our modeling results show that Chiplets are 4 times cheaper than SoCs.

\subsubsection{{{Power}}}
Power is modelled as follows
\begin{equation}
pSwitch=A*C*F*V^2
\end{equation}
where A is switching frequency, C is load capacitance, F is frequency, V is input Voltage
\begin{equation}
pShort=A*(B/12)*F*T*(V-2*Vth)^3
\end{equation}
where, B is gain factor , T is rise/fall time of gate and Vth is threshold voltage.
\begin{equation}     
pLeakagePerTransistor=I*V
\end{equation}
Where, I is leakage current.
\begin{equation}
pLeakage= pLeakagePerTransistor*TDensity*Area
\end{equation}
Where, TDensity is number of transistors per mm2.
\begin{equation}
pTotal=pSwitch+pShort+pLeakage
\end{equation}
We do Dynamic Volatge and Frequency Switching for the chiplet and assign the hyper parameters for each tile as shown in the code below. Results show that power consumed by a chiplet is roughly the same as power consumed by Soc.

\subsubsection{{{Performance}}}
Performance depends on two factors, throughput and latency. Latency is calculated as follows:
\begin{equation}
Latency= bi/Ri+Ti
\end{equation}
where, bi is represents what bit processor the core is, Ri(service bandwidth) is directly proportional to frequency and Ti is the maximum latency of the service. The throughput is directly proportional to frequency multiplied by the number of data channels. Since we assume that an SoC can be decomposed into 4 chiplets, we can assume that the number of channels in a Chiplet is atleast 4 times the number of channels in an SoC.

We calculate performance ratio of SoC vs Chiplet by taking ratio of throughputs and latency. Results show that chiplets are 4.5X better in performance than SoCs.

Hence, on optimizing cost, power and performance, Chiplets seem to be clear winners.

\section{Application of Chiplets in Automotive Compute Stack} 
In automotive computing, Chiplets are gaining significant attention and prominence due to several key factors, including
\begin{itemize}
\item The increasing demand for the integration of dissimilar Si-technologies and IP blocks within the automotive package to meet the rigorous requirements of modern vehicle electronics. 
\item The provision of interconnect and 3D stacking technologies that offer low power, low latency, and high-bandwidth connectivity to deliver System in Package (SIP) performance comparable to a monolithic die.
Performance comaparision was analyzed quantitatively in chapter 3, under the section “Analysis of Chiplet”, in this document.
\item Chiplet technology enables shorter and simpler IC design cycles by using pre-existing chiplets, reduces dependency on leading-edge fabrication, and improves time-to-market and agility in the fast-changing automotive sector.
\item Small die size enhances yield resiliency, which is vital for advanced silicon technology nodes in automotive applications.
\item Flexibility in picking the best process node for the IP—especially for SerDes I/O, RF, and analog IP that do not need to be on the “core” process node. 
\item Chiplets enable cost savings by reusing the same functional blocks across different designs, integrating them in a modular fashion, and buying only known-good dies(KGD).
\end{itemize}
\cite{b26}\cite{b25}\cite{b24}

Here are some sectors that can benefit greatly from using chiplets.

\subsection{Advanced Driver Assistance Systems \& Autonomous driving}
ADAS/AV processors demand advanced Si nodes, high-performance computing capabilities, AI/ML algorithms, high-speed SERDES IOs, expanded memory integration and all of this in a Compact footprint. These needs are mirrored in AI/ML packaging solutions similarly driven by processing speed, low latency, bandwidth. Heterogeneous packaging technologies empower product architects to optimize performance, reduce costs, and enhance time-to-market. Future vehicles, resembling mobile servers, require integration of diverse silicon products using 2.5D/3D packaging, with mass adoption expected over 5-10 years, especially in mainstream segments. To make autonomous driving accessible across price points, cost-effective, high-performance SoCs leveraging chiplet architectures and advanced packaging are crucial.

\subsection{Infotainment}
Consumers expect a seamless in-vehicle experience, driven by the technology pace in personal devices.While satisfying the requirements stated in Section \ref{A}, requires necessitating advanced hardware and software.

The use of chiplets in the infotainment system helps in reducing the cost of ownership, enabling the separation of safety from non-safety related applications, and driving virtualization down to the hardware. Chiplets are a promising approach to enable the open platforms for the software-defined digital cockpit, and a new generation of high-performance, heterogeneous SoCs is required to support the range of applications in the increasingly complex cross-domain digital cockpit.

\section{Special Purpose Computing for Automotive Applications} 
\subsection{Introduction to Neuromorphic Computing}
The traditional computation systems today use von Neumann architecture,( the system has three main components—a processor, memory, and storage). The industry, however, is running into an I/O bottleneck with today’s systems along with a lot of time and energy being spent in shipping the data around. 

For these applications, the industry hopes to develop a new class of neuromorphic systems where the memory and chip are processed together. This enables faster processing, unlike the current chips based on SRAM as they are power-hungry and have a large footprint. Hence neiromorphic computing utilises next-generation memory types, such as phase-change or ReRAM. In addition, phase-change and ReRAM promise to enable spike-based learning techniques in chips that are closer to biological neural networks where information is transmitted in the form of spikes (sNN: spike neural networks). 

\subsection{Applications in Automotive Domain}

There is no doubt about the growing application of AI Systems and Machine Learning Algorithms in the automotive space ranging from purposes like Infotainment, Data Analysis, and Safety systems built on data (Anomaly Detection) to ADAS, Autonomous Driving (going upto Level 5 Autonomy). The increasing complexity and dimensions of ML models have led to an exponential spike in the latency, power, large operational resources, footprint, and data storage required from the hardware. Even as the technology nodes continue to scale down, the total fabrication costs and design challenges associated with large-scale neural network hardware accelerators keep on growing as well. Additionally, the total development time is increased due to the design complexity and fabrication yield issues. Neuromorphic computing aims to solve this pivotal issue.

Neuromorphic Computing utilizes semiconductors to mimic neuro-biological architectures to drive adaptive computing. This greatly improves the speed, programmability, and capacity of neuromorphic processing, broadening its usage in power and latency-constrained intelligent computing applications. It delivers orders of magnitude improvements in energy efficiency, speed of computation, and efficiency of learning across a range of edge applications: from vision, voice, and gesture recognition to search retrieval, robotics, and constrained optimization problems. spike-based neuromorphic hardware provides more energy-efficient implementations of Deep Neural Networks (DNNs) than standard hardware such as GPUs.

\begin{table}[h!]
\centering
\begin{tabular}{lllll}
                       & \multicolumn{4}{c}{ImageNet and LSUN Dataset}     \\
                       & Energy (kW/h) & Saving & Time (h) & Saving \\
Neuromorphic computing & 0.51         & -      & 6.3     & -      \\
GPU                    & 3.1          & 6.1x   & 17      & 2.7x   \\
FPGA                   & 0.79         & 1.5x   & 30      & 4.8x   \\[1ex]
                       & \multicolumn{4}{c}{LSUN Dataset}                  \\
                       & Energy (kW/h) & Saving & Time (h) & Saving \\
Neuromorphic computing & 3.8          & -      & 47.2    & -      \\
GPU                    & 23.4         & 6.1x   & 130     & 2.8x   \\
FPGA                   & 5.5          & 1.4x   & 255     & 5.5x   
\end{tabular}
\caption{Comparison of Energy and Time Savings Across Different Datasets and Computing Architectures}
\end{table}

\section{Overview of Communication Solutions} 

Communication technologies, in a broader context, encompass a diverse array of tools and techniques designed for the reception, transmission, and exchange of information between two entities.

In the realm of System-on-Chip (SoC) architecture, the challenge lies in integrating various smaller components within a monolithic die. These components need to seamlessly communicate and synchronize data among themselves, typically under the guidance of a shared clock signal.

When considering the concept of chiplets, which involves breaking down a monolithic chip into smaller, more manageable pieces, the imperative remains the same. Chiplets necessitate the exchange of information among themselves in a synchronized and harmonious manner. This synchronization is essential for the efficient functioning of chiplet-based systems.

\subsection{Current Communication Technologies in SoC Architecture}

\subsubsection{Bus based technologies}
Serial and parallel buses are some of the earliest answers to the on-chip communication problem. These involve an initiator, a target and a bus arbiter, like other standard bus protocols. Chip makers developed and owned their IP standards for interconnects. One of the earliest open standards in on-chip bus communication is the ARM-developed Advanced Micro-Controller Bus Architecture (AMBA). The Advanced eXtensible Interface (AXI) technology developed on AMBA standard is one of the most used and developed interconnect solutions across chip makers. Avalon Bus, Wishbone, Sonic SMART Interconnects, STBus and CoreConnect Bus are other primary bus protocols used in the industry. Tradional buses like SPI, I2C and CAN also have on-chip implementations, offering increased bandwidth and reduced power consumption. \cite{b13}

\begin{itemize}
\item Advanced eXtensible Interface (AXI): AXI is a high-speed protocol for Systems-on-Chip. It operates with separate read and write channels for simultaneous data transfer. AXI supports advanced features like burst-based transactions, out-of-order data transfer, and multiple outstanding transactions, ideal for complex SoC architectures. Its versatility and support for high-bandwidth data exchange make it a preferred choice for sophisticated and high-performance system designs.
\item Avalon Bus: Avalon is a bus standard developed by Intel for FPGA design IP cores. It employs a structured protocol, enabling diverse components to communicate within the FPGA architecture, supporting various functionalities and signal types for seamless integration.

\item Wishbone: Wishbone is a simple, open-source, and widely used on-chip bus architecture specifically designed for connecting various components within a System-on-a-Chip (SoC). It’s a popular choice for connecting slower peripherals due to its ease of implementation and low power consumption.
\end{itemize}

\subsubsection{NoC - Network-on-Chip}
A network-on-chip (NoC) is a communication infrastructure that connects different components within a system-on-chip (SoC). SoCs are integrated circuits that combine multiple processors, memory blocks, and other components onto a single chip. NoCs provide a high-bandwidth, low-latency communication fabric that allows these components to communicate with each other efficiently.
NoCs are typically designed using a mesh topology, which means that each component is connected to multiple other components. This topology provides redundancy and helps to ensure that data can be routed around any failed components. NoCs are becoming increasingly important as SoCs become more complex. By providing a scalable and efficient communication infrastructure, NoCs enable SoC designers to create more powerful and versatile chips.

\subsubsection{SerDes}
SerDes (serializer/deserializer) is a high-speed communication technology used to serialize and deserialize data for transmission over long distances. SerDes are used in a wide range of semiconductor designs, including processors, memory, and graphics chips. SerDes are important in the semiconductor industry because they enable high-speed communication between different chips and devices. This is essential for a wide range of applications, including networking, data centers, and high-performance computing. SerDes offer a number of benefits over other communication technologies, including:

\begin{itemize}
    \item High bandwidth: SerDes can support very high bandwidths, up to 100 Tbps and beyond.
    \item Low latency: SerDes have very low latency, making them ideal for applications where real-time communication is required.
    \item Power efficiency: SerDes are very power efficient, making them ideal for battery-powered devices.
    Scalability: SerDes can be scaled to support a large number of devices and connections.
\end{itemize}

SerDes technology is constantly evolving to support higher bandwidths, lower latencies, and lower power consumption. This is being driven by the increasing demands of applications such as networking, data centers, and HPC. In the future, SerDes are expected to play an even more important role in the semiconductor industry as the industry moves towards chiplet-based designs and 3D-ICs.

\subsubsection{Latest developments in on-chip communication}

In recent years, advanced technologies with niche applications have been developed.

\begin{itemize}
\item Automotive busus: These on-chip bus protocols facilitate the exchange of data and commands among different modules and components within the automotive electronic systems. AMBA AXI, Avalon Bus and Xilinx AXI4 are on-chip implementations suited for automotive applications.

\item Optical interconnects: Optical interconnects solve the wiring issue in buses by employing light-based signals instead of traditional electrical signals. This enables high-speed, low-latency, and energy-efficient data transmission, addressing the limitations of traditional copper-based interconnects. Optical interconnects utilize photonics to integrate lasers, modulators, and detectors, directly onto the chip. These interconnects offer increased bandwidth, reduced heat dissipation, and potentially improved performance, making them a promising solution for future high-performance computing.
\item Plasmonic interconnects: The use of Plasmonics in SoC interconnects is very new and an area of active research. They use the collective oscillations of electrons in a metal when excited by light, for data transmission within and between integrated circuits. Such communication technologies have the potential to achieve high data transfer rates, overcome limitations related to signal interference and latency. 
\end{itemize}

\subsection{Current Communication Technologies in Chiplet Architecture}
\subsubsection{AIB - Advanced Interface Bus}

AIB\cite{b14} is parallel bus architecture developed by Intel which was primarily developed for high bandwidth applications of about 2 Gbps that can support a wide range of data types, including signals from sensors, actuators, and memory. It had a much simpler architecture in comparison to PCIe. One of the key features of AIB is its use of a microbump interface. Microbumps are small solder bumps that are used to connect chips or chiplets together. The microbump interface allows AIB to achieve high-speed data transfer with low latency.
\newline
The AIB interface has many input/output (I/O) signals. These signals can be used to transmit (TX) or receive data (RX). The I/O signals are grouped together into channels, which are then grouped together into columns with each column having upto 24 channels. Each channel can be configured to be used only for transmitting data, only for receiving data, or both.
\newline
AIB is an architecture which has I/Os equipped with active redundancy. Multiple spare microbumps are present in case there is a microbump-fault. In case of a fault, the I/O signals shift over to make use of neighbouring microbumps while utilising the spare microbumps which ensures higher reliability and manufacturing yield.
\newline
The architecture makes use of a master-slave configuration wherein The master device initiates all of the data transfers and the slave devices respond to the master device's requests enabling high degree of control over the data flow.

There are a few problems associated with AIB include the effects of skew which leads to reduced data transfer rates and increased bit error rates. While making use of DDR (double data rates) it is important to maintain a duty cycle of 50\% with minimal error to ensure reliable data transfer. Special accommodations such as clock forwarding and duty cycle correction circuits are implemented to ensure that the aforementioned problems are mitigated. 
\subsubsection{UCIe - Universal Chiplet Interconnect Express}

\cite{b15}

UCIe adopts a comparable strategy to the widely recognized Peripheral Component Interconnect Express (PCIe), a standardized interface on printed circuit boards that permits the integration of diverse devices for functions like graphics, memory, and storage. UCIe extends this concept to the micro-level, facilitating interconnects between individual semiconductor dies. Notably, UCIe garners support from prominent industry leaders, including AMD, Arm, ASE, Google, Intel, Meta, Microsoft, Qualcomm, Samsung, and TSMC. 

When we compare UCIe to PCIe, we observe substantial enhancements. UCIe's linear bandwidth on the shoreline ranges from 28 to 224 in a standard package and 165 to 1317 GB/s/mm in an advanced package, representing an improvement of over 20 to more than 100 times. In terms of latency, PCIe typically operates at around 20ns, while UCIe achieves less than 2ns (for both transmission and reception), marking a tenfold improvement. Moreover, the power efficiency of UCIe stands at 0.5 pJ/b for the standard package and 0.25 pJ/b for the advanced package, signifying a more than tenfold advancement. This increased power efficiency not only results in reduced heat generation but also enhances the overall reliability of semiconductor devices, making these improvements highly significant.

UCIe, with its initial version designed for both 2D and 2.5D processes, is joined by a forthcoming UCIe 3D process in development. This 3D process aims to simplify chiplet connections and address some of the current manufacturing challenges.

UCIe 1.0 stands as the inaugural open industry standard to provide backing for the die-to-die I/O physical layer, die-to-die protocols, and software stack, all rooted in the industry standards of PCI Express (PCIe) and Compute Express Link (CXL).

The UCIe specification, as depicted in the above Figure, is structured into three distinct stack layers:

Physical Layer: This layer serves as the electrical interface to the package media. It encompasses components such as the electrical AFE (transmitter and receiver) and a sideband channel that facilitates parameter exchange and negotiation between two dies. Additionally, it includes the logic PHY responsible for implementing link initialization, training, calibration algorithms, as well as test and repair functionality.

Die-to-Die Adapter Layer: This layer manages link functionality and handles protocol arbitration and negotiation. It also incorporates the option for error correction, which is based on a CRC (Cyclic Redundancy Check) and retry mechanism.

Protocol Layer: The Protocol Layer is responsible for implementing one or more of the UCIe-supported protocols. Currently, these protocols include PCI Express, CXL, and/or streaming, all of which are Flit-based protocols, designed to optimize efficiency and minimize latency.

In UCIe's terminology, a module is established as the most compact interface unit. Each module comprises a primary "bus" with a capacity of up to 64 transmit and receive IOs for advanced packages (or 16 for standard packages), in addition to clock forward IOs, valid (framing), and track IOs. Additionally, a sideband "bus" is incorporated, as illustrated in the above figure.

In advanced package assembly, UCIe introduces a test and repair mechanism to minimize yield losses stemming from poor bump quality. This mechanism relies on six redundant pins for TX and RX data, clock, valid, and track, along with two redundant pins for sideband TX and RX.

For standard packages, UCIe takes a different approach, as C4 (or CuPillar) bump yield and the entire assembly process yield are typically high. In these instances, UCIe supports a "degraded" operating mode. This mode activates only half of the module if a failure is detected on the other half.

The test and repair procedure occurs during link initialization. The PHY component evaluates each die connection to identify any faults. If a failure is detected, the corresponding signal is re-routed to a redundant pin.

\subsubsection{BoW - Bunch of Wires}
\cite{b16}

BoW is a chiplet interconnect specification that is versatile, open, and interoperable. It is energy-efficient and easy to use, and it is designed to connect die placed close to one another within the same package. BoW uses a lower data rate per wire, which requires more wires than SerDes, but it allows for single-ended signaling and denser wire packing. BoW can also take advantage of multiple wiring layers in laminates and the increased wire density in advanced packaging.

\begin{figure}[htbp]
    \centering
    \includegraphics[width=0.6\textwidth]{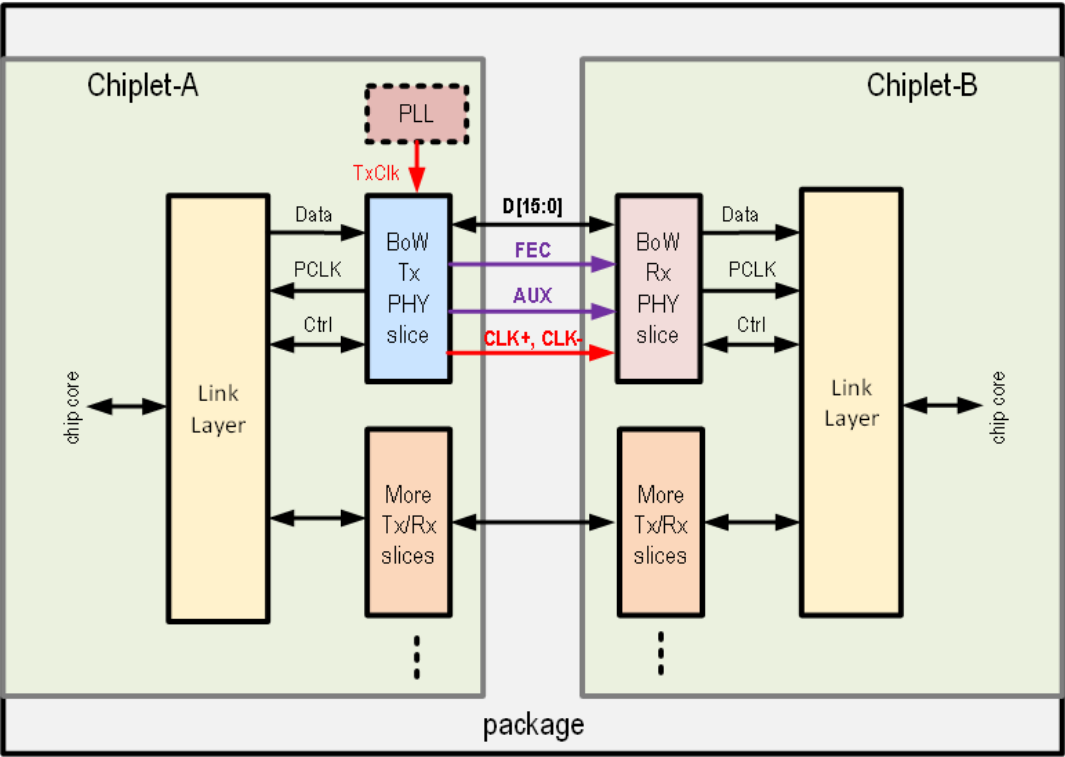}
    \caption{Block Diagram for BoW Architecture}
\end{figure}
BoW is defined as a single unidirectional slice, and multiple slices can be used to create a bidirectional interface. BoW is fully backward compatible with BoW 1.0, and it is flexible to support both laminate and advancing packaging technologies. BoW is also portable across IC process nodes ranging from 65 nm to 5 nm and beyond.

BoW modes are defined by the speed of clock and data of the PHY on the die. The slice data rate is the rate at which data is transmitted on a single slice, while the wire bit rate is the rate at which data is transmitted on a single wire. TX Clk is the clock frequency of the transmitter. The table \cite{b17} shows the following BoW modes: 

\begin{figure}[htbp]
  \centering
  \begin{minipage}[b]{0.5\textwidth}
    \centering
    \includegraphics[width=\textwidth]{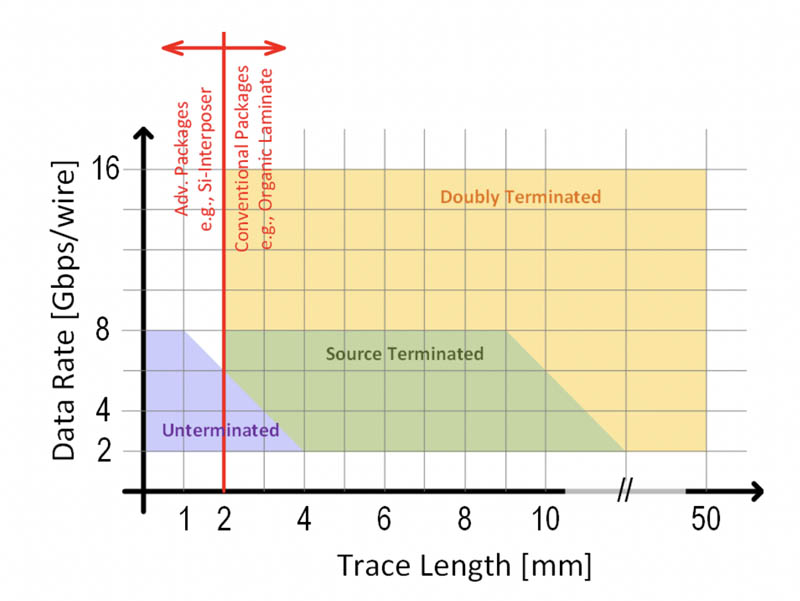}
    \caption{BoW Trace Length And Data Rate}
  \end{minipage}%
  \begin{minipage}[b]{0.5\textwidth}
    \centering
    \includegraphics[width=\textwidth]{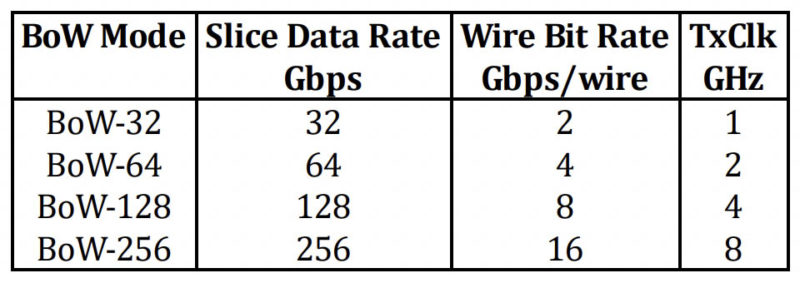}
    \caption{BoW Speeds}
  \end{minipage}
\end{figure}

\subsubsection{QPI - QuickPath Interconnect and UPI - UltraPath Interconnect}
QPI is a high-speed, packet-based interconnect that uses a serial link to connect two processors but is no longer in active usage. It can support data rates of up to 6.4 GT/s per direction, for a total bandwidth of up to 25.6 GB/s. The interconnect ports are connected together in a ring architecture. In a ring architecture, all of the components in the system are connected in a loop. This means that data has to travel through all of the components in the ring in order to reach its destination. This can lead to high latency and bandwidth constraints, especially as the number of components in the system increases.

UPI which also makes use of a serial link was developed to improve the data rate and power efficiency primarily for mesh architectures. It supports a data rate upto 10.4 GT/s per direction and a net bandwidth of 41.6 GB/s. 
In a mesh architecture, each component is connected to multiple other components. This allows data to take the shortest path to its destination, which can significantly improve latency and bandwidth. Such architectures are much more flexible and scalable compared to ring based interconnect architectures. 

\begin{figure}[htbp]
  \centering
  \begin{minipage}[b]{0.5\textwidth}
    \centering
    \includegraphics[width=\textwidth]{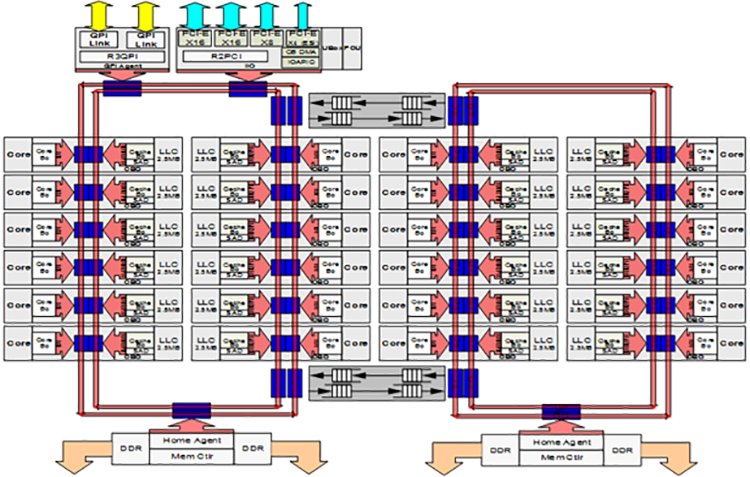}
    \caption{Ring-based interconnect architecture}
  \end{minipage}%
  \begin{minipage}[b]{0.5\textwidth}
    \centering
    \includegraphics[width=\textwidth]{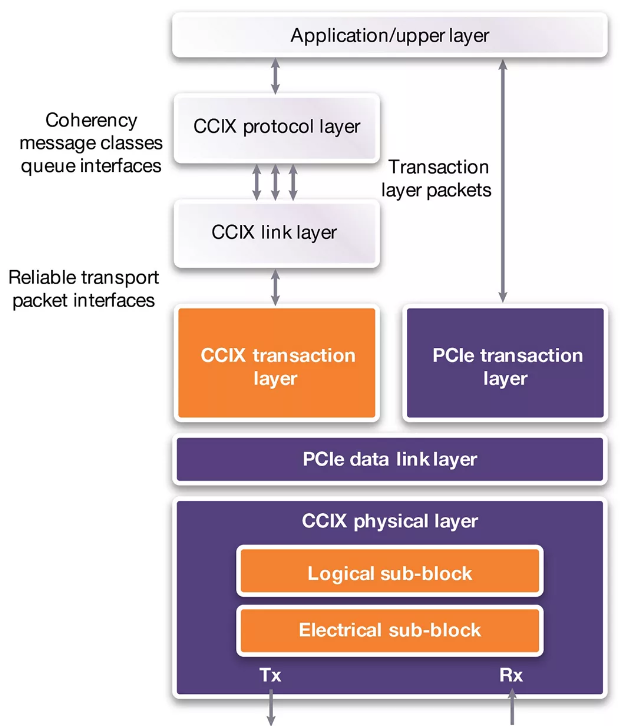}
    \caption{CCIX Architecture}
    \label{fig:immagine}
  \end{minipage}
\end{figure}

 \subsubsection{LIPINCON} 
 LIPINCON (Low-voltage-In-Package-INterCONnect) is a parallel-bus chiplet interconnect architecture developed by TSMC, designed for advanced packaging technologies such as InFO (Integrated Fan-Out) and CoWoS (Chip-on-Wafer-on-Substrate). It offers data rates of upto 8Gb/s per pin. It also has a power efficiency 0.56 pJ/bit and a bandwidth of  1.6$Tb/s/mm^{2}$.
 \newline
It features a clock-forwarded architecture that uses a timing compensation mechanism to achieve low-power consumption. Such an interconnect can support various workloads including machine learning applications. 
\subsubsection{CCIX}
Cache Coherent Interconnect for Accelerators (CCIX) was proposed with the main focus being on faster interconnects as well as incorporating cache coherency which allows for faster memory access in heterogeneous multi-processor systems. 
\newline

CCIX is built on PCIe, thus enabling any existing PCIe controller implementation to be used to implement a CCIX transaction layer.  
It was determined that to achieve 100 GB/s of data transfer a feature called Extended Speed Mode (ESM) was established to tackle the bottleneck of PCIe having the highest data rate specification of 16 GB/s. Depending on the component's (in this case chiplet) data transfer capability, the standard PCIe mode or ESM mode of data transfer could be chosen. 

\subsubsection{OpenCAPI}
Open Coherent Accelerator Processor Interface (OpenCAPI) is an open industry standard device interface. OpenCAPI utilizes high-frequency differential signaling technology that provides high data bandwidth with low latency which is required by the emerging advanced memory technologies. 
\newline
The need for coherent high performance bus arises from the requirement of hardware acceleration. IBM's POWER9 microprocessor achieved upto 25 Gb/s data rate with very low latency. Future works promise upto 56 Gb/s rate of data transfer. In comparison to PCIe, it provides much less power consumption and is much more scalable as well as flexible. 

\subsubsection{Gen-Z}
Gen-Z makes use of specialised data Movers for CPU offload which enables high-performance, reliable, and secure data communication in interconnect systems. It also has a very sophisticated arbitration system for handling data traffic while also being equipped with multiple CRC codes to protect against single-bit, multi-bit and burst errors. Gen-Z can utilize both electrical and optical interfaces, Optical interconnects are advantageous for applications that demand higher bandwidth and longer reach. Optical interconnects can offer upto 1 Tbps/optical module with 6 fibers per direction. 

\subsubsection{CXL}
Compute Express Link (CXL) is an open standard which is built on serial PCIe physical and electrical interface. It also makes use of cache coherent protocols for access to  system and device memory. The CXL standard supports a variety of applications via three protocols: CXL.io, CXL.cache, and CXL.memory. CXL memory controllers generally introduce a latency of about 100 to 200 nanoseconds while it has a data transfer of 64 Gb/s per x16 linkage. 
\begin{figure}[htbp]
    \centering
    \includegraphics[width=0.8\textwidth]{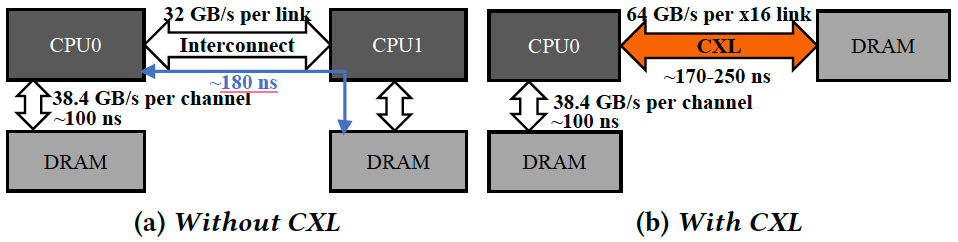} 
    \caption{Comparison of latency of system with and without CXL} 
    \label{fig:immagine}
\end{figure}
\subsection{SUMMARY}

\begin{table}[ht]
\centering
\caption{Interconnect specifications}
\label{table:interconnect1}
\setlength{\tabcolsep}{1pt} 
\begin{tabular}{@{}c c c c c c@{}}
\toprule
\textbf{Interconnect} & \textbf{Bandwith } & \textbf{Latency } & \textbf{Energy Eff. } & \textbf{Channels} & \textbf{Bump density} \\
\textbf{} & \textbf{ (GB/s/mm)} & \textbf{ (ns)} & \textbf{(pJ/b)} & \textbf{} & \textbf{ (micron)} \\
\midrule
UCIe & 28 to 1317 & 2 & 0.25 to 0.5 & 16 to 64 & 25 to 55 \\
AIB & 256 to 1024 & 3.56 & 0.85 & 40 to 160 & 55 \\
BoW & 1280 & 5 & 0.5 to 1 & 16 & 40 \\
QPI & 6.4GT/s & 100 & 10 to 40 & 20 & 400 \\
CXL & 32GT/s & 100 to 200 & 6 & 32 & - \\
Infinity fabric & 5000MT/s & 200 & - & 32 &  - \\
LIPINCON & 1.6Tb/s/$mm^2$ & - & 0.56 & - &  -\\
\bottomrule
\end{tabular}
\end{table}

\section{IPs required for Communication}
\subsection{UCIe PHY IP for TSMC N5 by Synopsys}

Synopsys UCIe PHY IP is specifically designed to cater to the demands of hyperscale data centers, AI applications, and networking scenarios. It is a versatile and high-performance solution that provides a range of benefits for these critical domains.

One of the standout features of this PHY IP is its exceptional bandwidth and flexibility. It not only supports both standard and advanced packaging technologies but also boasts an impressive bandwidth of up to 4 Tbps in a multi-module configuration. This capability is crucial for data-intensive applications, ensuring seamless data transmission and processing.

\begin{figure}[htbp]
    \centering
    \includegraphics[width=0.6\textwidth]{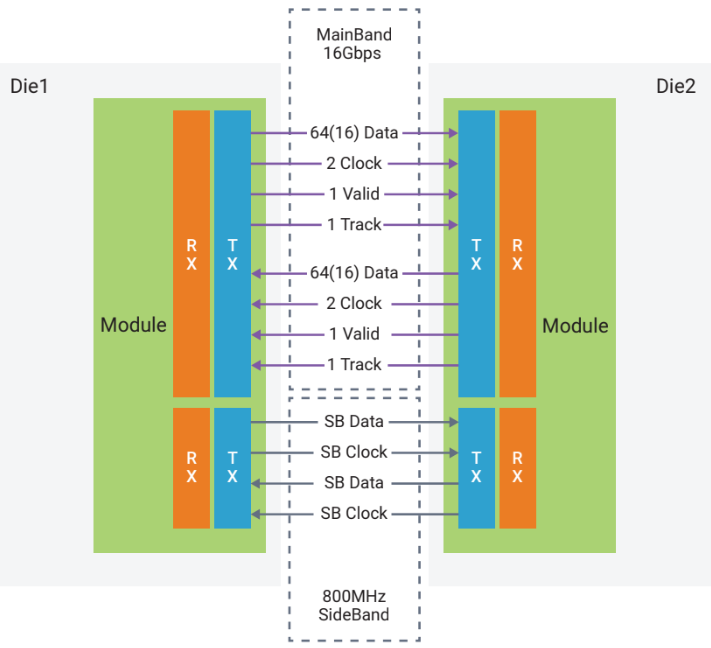} 
    \caption{Synopsys UCIe PHY IP architecture (one module)} 
\end{figure}

Furthermore, the UCIe PHY IP is engineered to offer extensive protocol support. It seamlessly integrates with widely used protocols such as PCI Express and CXL, enabling low-latency NoC-to-NoC links with streaming protocols. This ensures that it can effectively communicate and transfer data in a standardized and efficient manner, vital for compatibility and ease of integration.

In terms of performance, the IP delivers maximum performance with minimal latency, giving it a competitive edge in data center and AI environments where speed and responsiveness are paramount. Its implementation flexibility further enhances its adaptability to diverse application requirements.

Energy efficiency is another key aspect of this PHY IP. It prioritizes high energy efficiency through features like clock forwarding and low-voltage signaling. This not only contributes to energy savings but also supports sustainable operations, an increasingly important aspect of modern technology solutions.

Testability is also a core focus of the UCIe PHY IP. It incorporates a comprehensive set of testability features for known good dies, offering robust test and repair capabilities to enhance package assembly yield. This is crucial for ensuring product quality and reliability.

The IP's robust operation is guaranteed through embedded training and calibration algorithms, which ensure reliable die-to-die link operation. This robustness is essential in demanding, high-performance applications where downtime or errors are not acceptable.

Finally, the interoperability of the Synopsys UCIe PHY IP is a significant advantage. It seamlessly integrates with the Synopsys UCIe Controller IP, forming a complete, low-latency solution for die-to-die links in any package. This interoperability enhances the overall efficiency and ease of implementation of the technology.

\subsection{112G XSR IP by Synopsys}
The Synopsys XSR PHY IP for 112Gbps per lane die-to-die connectivity is a cutting-edge solution designed to cater to the high-bandwidth demands of multi-chip modules (MCMs) in hyperscale data centers, AI applications, and networking environments. This IP offers low-latency, low-power, and compact physical layer support for data rates ranging from 2.5G to 112G, making it suitable for a wide range of applications.

One of the standout features of the XSR PHY IP is its flexibility in supporting NRZ and PAM-4 signaling, catering to various data rate requirements. It complies with OIF CEI-112G and CEI-56G standards for extra-short reach (XSR) links, ensuring interoperability and standard compliance. The PHY allows for flexible layout options, maximizing bandwidth per die-edge, and supports up to 16-lane transmit and receive macros, optimized for segmentation across multiple dies.

\begin{figure}[htbp]
  \centering
  \begin{minipage}[b]{0.5\textwidth}
    \centering
    \includegraphics[width=\textwidth]{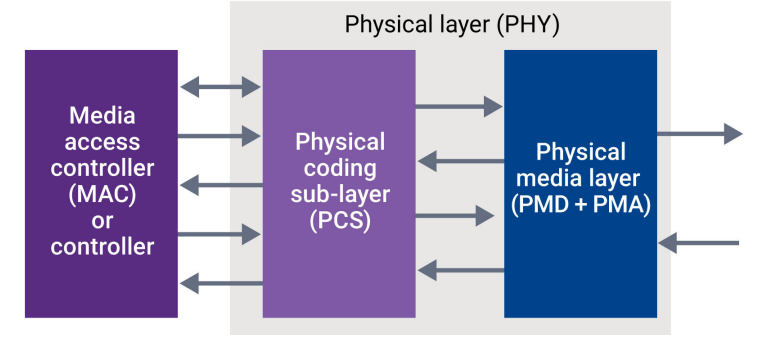}
    \caption{Synopsys XSR PHY IP}
  \end{minipage}%
  \begin{minipage}[b]{0.5\textwidth}
    \centering
    \includegraphics[width=\textwidth]{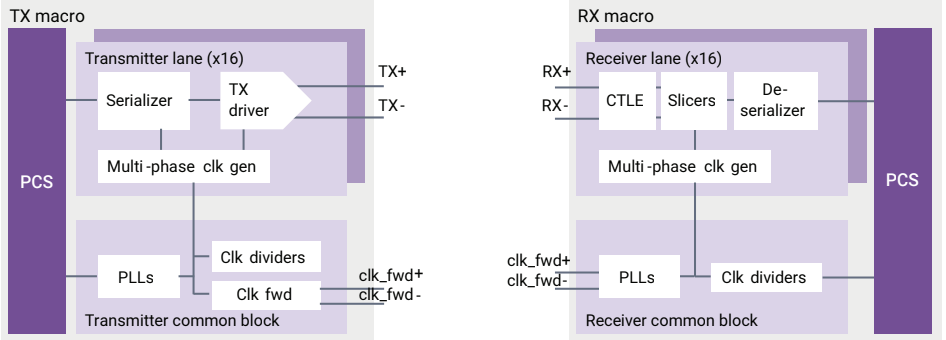}
    \caption{Synopsys XSR RX and TX IP block diagram}
  \end{minipage}
\end{figure}

Efficiency and energy savings are paramount in this IP solution. It employs a robust DLL-based clock forwarded architecture, which not only minimizes complexity but also supports reliable links of up to 50 millimeters for large MCMs. This feature enables multi-die connectivity over organic substrates, reducing packaging costs and eliminating the need for advanced interposer-based packaging over shorter distances.

In terms of testability, the XSR PHY IP incorporates an embedded bit error rate (BER) tester and non-destructive 2D eye monitor capability, providing on-chip testing and visibility into channel performance. It also includes a raw-PCS to facilitate interfacing with on-chip networks, regardless of the existing networking protocol. This comprehensive approach ensures thorough testing and integration capabilities.

Furthermore, the XSR IP seamlessly integrates with Synopsys' routing feasibility analysis, packages substrate guidelines, signal and power integrity models, and crosstalk analysis, facilitating fast and reliable integration into System-on-Chip (SoC) designs.

Key features of the Synopsys XSR PHY IP include 16 lanes of NRZ and PAM-4 transmitters or receivers, compliance with OIF XSR standards, clock forwarding, and embedded clock recovery algorithms, and shared PLLs for maximum energy efficiency. These features enhance performance, reduce design constraints, and ensure robust operation.

\subsection{AIB2.0 PHY IP by Synopsys}
The AIB2.0 PHY IP offers upto 6.4 Gbps in comparison to its earlier version AIB1.0 which could offer upto 2 Gbps. It also uses a much lower IO Voltage Output Swing of 0.4V which leads to reduction in power consumption. It functions with low latency in AIB-based multi-die systems. 

\subsection{Innolink IP}
The INNOLINK IP is based on a novel architecture that uses standard organic substrates, eliminating the need for complex silicon interposers. It is compatible with the UCIe standard.
\newline
It incorporates incorporates DDR-like protocols in its development. Three variants of the PHY IP were made available - Innolink A, B, C. Innosilicon adopted DDR-like protocols in Innolink-B and C PHY development to provide a high-speed, high-density, and high-bandwidth interconnection solution.
\subsection{Eliyan IP}
Eliyan’s 40Gbps/bump chiplet interconnect silicon implemented in a 5nm standard demonstrates the capability to achieve data bandwidth density up to 3 Tbps/mm. Eliyan's chiplet interconnect technology has been the foundation of the Bunch of Wires (BoW) standard and is fundamentally compatible with the UCIe standardization efforts.

\subsection{AresCORE16 IP}

AresCORE 16G Die-to-Die IP is a low-power, low-latency interface IP designed by Alphawave Semi. It implements a clock forwarded physical interface for multichannel connections upto 16 Gbps per pin. It is possible to configure the PHY IP with industry standards such as BoW, UCIe, CoWoS and InFO. 

\begin{figure}[htbp]
  \centering
  \begin{minipage}[b]{0.5\textwidth}
    \centering
    \includegraphics[width=\textwidth]{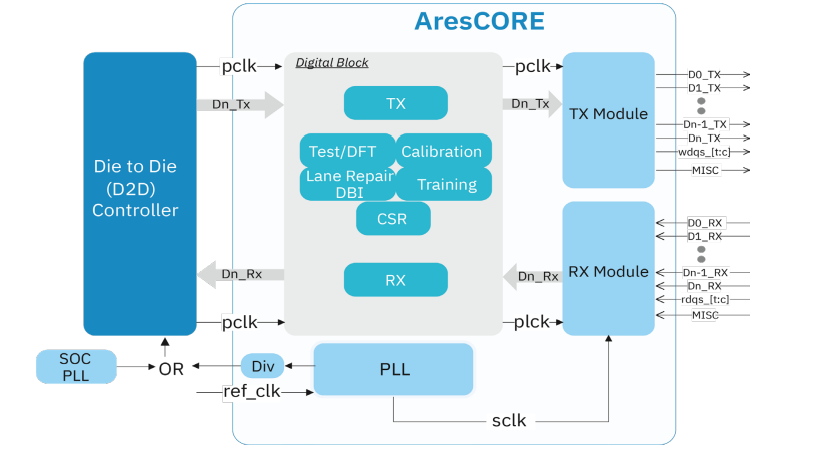}
    \caption{AresCORE 16G D2D IP Block Diagram}
    \label{fig:immagine}
  \end{minipage}%
  \begin{minipage}[b]{0.5\textwidth}
    \centering
    \includegraphics[width=\textwidth]{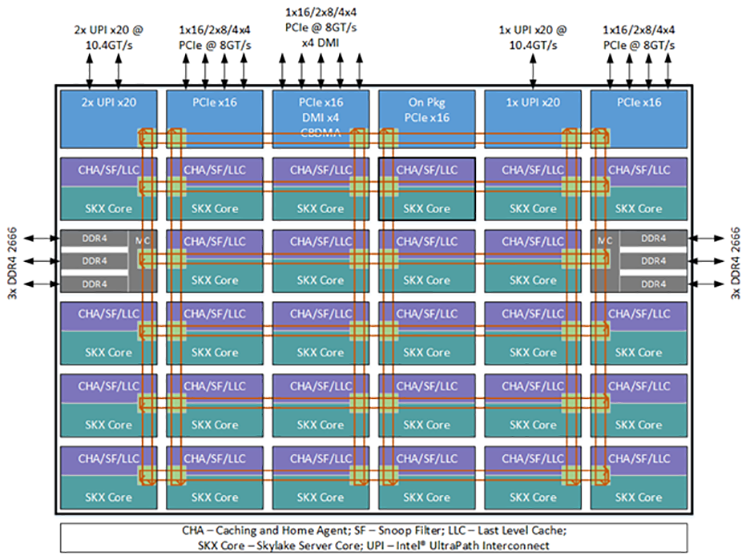}
    \caption{Mesh-based interconnect architecture}
  \end{minipage}
\end{figure}

The PHY IP also has a bandwidth density of 4 Tbps/mm while also having Built-In Self-Test Pattern (BIST) Generator. 

\subsection{Infinity Fabric}

AMD's Infinity Fabric is a proprietary interconnect technology integral to AMD's Ryzen and EPYC processors, known for its high-speed and low-latency capabilities. This advanced interconnect serves as the glue that connects different chiplets within the processor, including CPU chiplets, GPU chiplets, and memory chiplets.

Infinity Fabric is designed as a coherent high-performance fabric, utilizing embedded sensors within each die to facilitate efficient control and data flow from the die level to the socket and even to the board level. This level of control is vital for optimizing the performance and communication between various components in AMD's processors.

\begin{figure}[htbp]
  \centering
  \begin{minipage}[b]{0.5\textwidth}
    \centering
    \includegraphics[width=\textwidth]{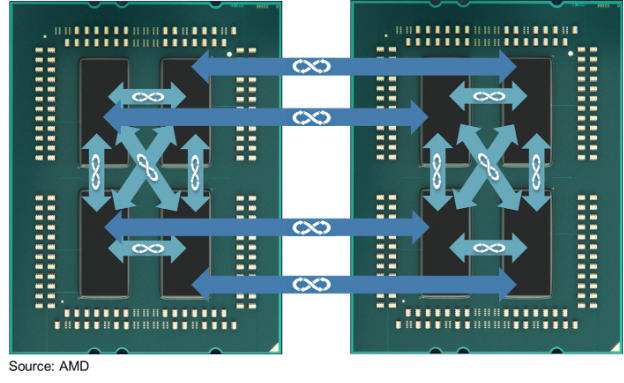}
    \caption{AMD Infinity Fabric connecting Zeppelin die on a MCM and between MCMs}
  \end{minipage}%
  \begin{minipage}[b]{0.5\textwidth}
    \centering
    \includegraphics[width=\textwidth]{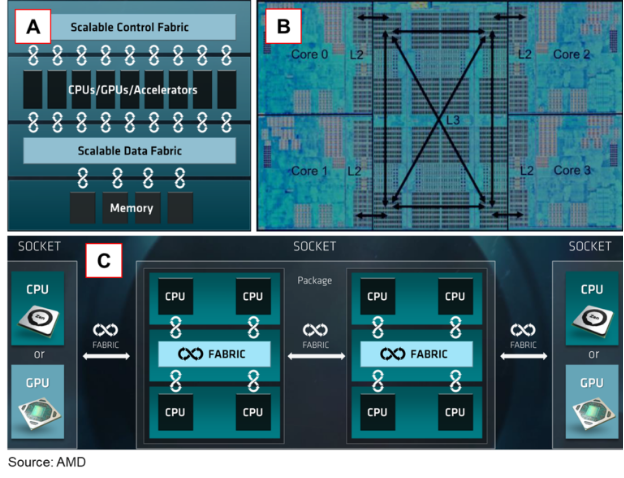}
    \caption{AMD Infinity Fabric high-level block diagram}
  \end{minipage}
\end{figure}

In terms of bandwidth, AMD measures the cross-sectional bandwidth of Infinity Fabric across the EPYC Multi-Chip Module (MCM) as four times that of 42.6 GB/s. This significant aggregate bandwidth of 41.4 GB/s per die in a two-socket system contributes significantly to the overall processor performance.

Moreover, AMD has specially tailored Infinity Fabric for two-socket servers, maximizing cross-sectional bandwidth to enhance its performance in such environments. This configuration includes a high lane count to facilitate efficient data flow.

Infinity Fabric is designed with interoperability in mind, ensuring compatibility with other technologies like PCIe 3.0. This means it can be utilized for various non-gaming, professional graphics rendering workloads in Epyc and Workstation Graphics systems.

One notable feature of Infinity Fabric is its System Control Fabric (SCF), which utilizes embedded sensors to monitor parameters like die temperature, speed, and voltage at a high rate of 1,000 times per second across all 64 cores in a dual-socket system. SCF, in conjunction with Epyc's system management unit, works in real-time to identify optimal operating settings. This level of control allows fine-grained frequency adjustments and customization of performance and power balance for specific workloads.

Infinity Fabric also governs power management, security, reset, initialization, and die test functions. If fewer cores are enabled, frequency boost is automatically activated while staying within temperature and current delivery limits.

The physical characteristics of the Infinity Fabric include 32 lanes in width between two Zeppelin dies, supporting full-speed operation on 16 lanes in each direction and operating at 5.3 Gb/sec for power conservation. The aggregate throughput between chips is 10.65 GB/sec point-to-point. The cross-sectional bandwidth of Infinity Fabric across the Epyc MCM is measured at 42.6 GB/sec, while the aggregate bandwidth on each die in a two-socket system reaches 41.4 GB/sec.

Infinity Fabric connections between sockets involve four connections, using the same SERDES links as 16 PCI-Express 3.0 lanes. These connections operate at approximately 9.5 Gb/sec each, providing a bidirectional bandwidth of 9.5 GB/sec between processors. In total, this results in an aggregate cross-section of 37.9 GB/sec bidirectional bandwidth between sockets.

Each Zeppelin die contains 256 dedicated Infinity Fabric and configurable I/O lanes, offering extensive connectivity options.

\begin{figure}[htbp]
  \centering
  \begin{minipage}[b]{0.5\textwidth}
    \centering
    \includegraphics[width=\textwidth]{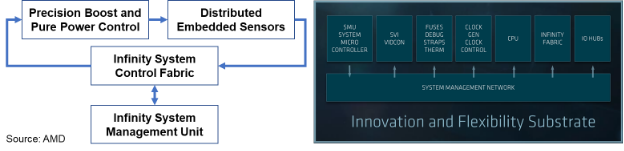}
    \caption{AMD System Control Fabric high-level block diagram and detail}
  \end{minipage}%
  \begin{minipage}[b]{0.5\textwidth}
    \centering
    \includegraphics[width=\textwidth]{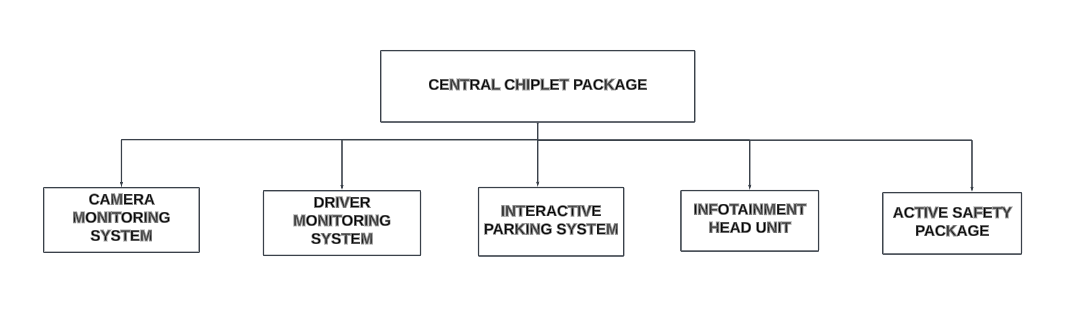}
    \caption{Domain split up}
  \end{minipage}
\end{figure}

Additionally, Infinity Fabric's Machine Check Architecture (MCA) is implemented with the technology, providing scalability and reporting MCA errors at both the die and link levels. It can implement data poisoning and reliability assurance measures to maintain the system's stability and performance.

In cases where a core becomes unreliable, potentially exceeding a threshold for deferred or correctable errors, a run-time operating system policy decision may be made to ignore it, allowing the remaining cores to continue operating effectively. This feature enhances the long-term reliability and performance of AMD's processors as the system ages.

\section{Micro-Architecture Diagram}
The initial stage in creating this micro-architecture diagram is to segment the car's requirements into domains and chiplet packages. Let us consider the following Advanced Driver Assistance systems and Infotainment unit [32][35][36][38].

This design framework hosts a centralised head and several sub-units of chiplet packages which might talk with each other through interconnects or through the central head. The placement of the above listed units in the vehicle is shown below.

\begin{figure}[htbp]
  \centering
  \begin{minipage}[b]{0.5\textwidth}
    \centering
    \includegraphics[width=\textwidth]{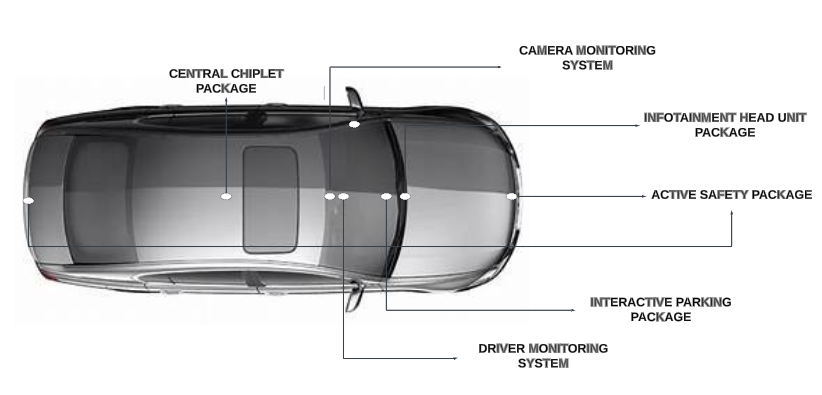}
    \caption{Placement of Various units}
  \end{minipage}%
  \begin{minipage}[b]{0.5\textwidth}
    \centering
    \includegraphics[width=\textwidth]{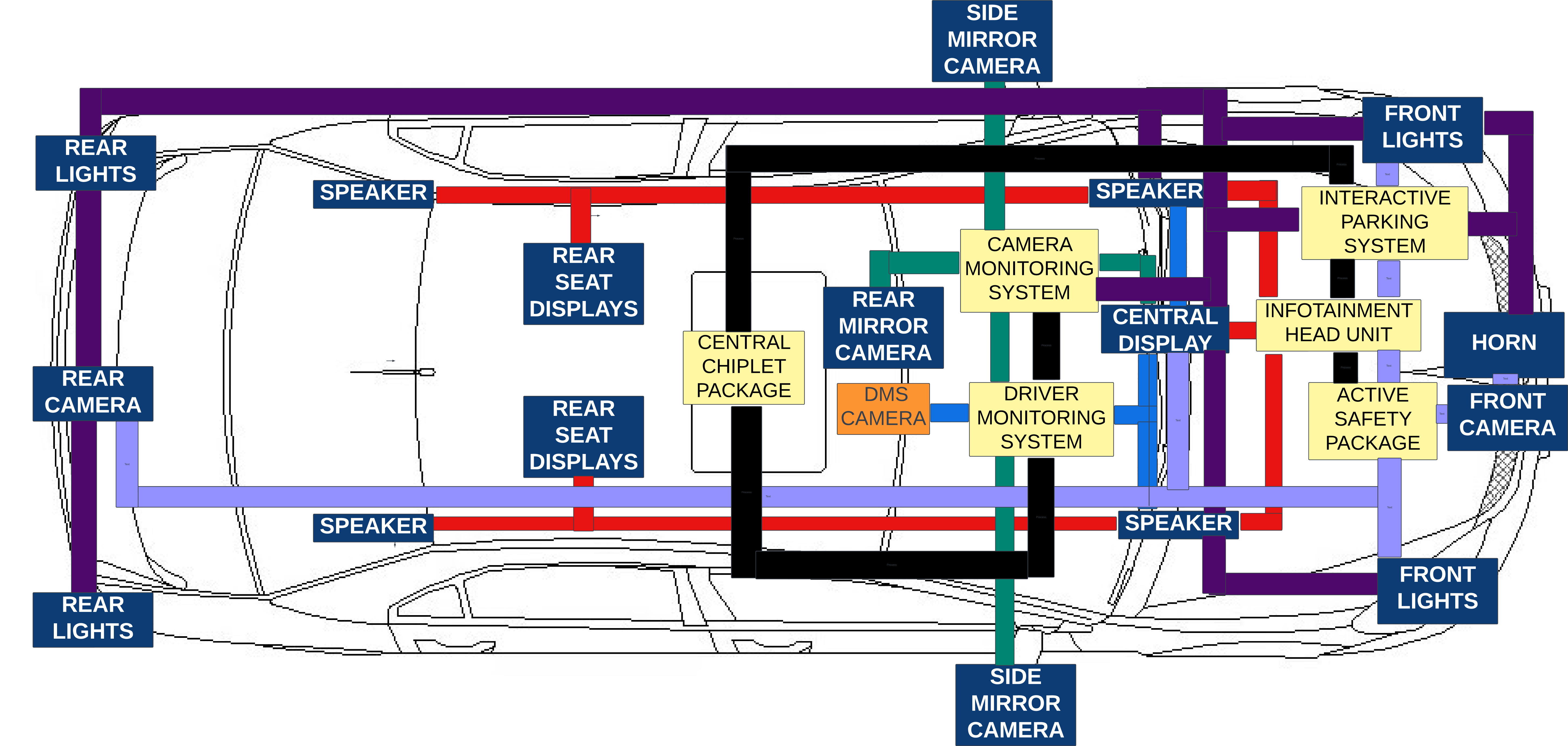}
    \caption{Interconnects between chiplets and peripherals}
  \end{minipage}
\end{figure}

\subsection{Design Specifications}
As mentioned in the chiplet architecture, each package of ADAS has been split into Compute Corelet, GPU Dielet, IO Dielet and Memory Dielet.

After conducting an extensive literature pore over of the state of art architectures in ADAS, we have concluded that the hardware requirements are close to the table given below [29][30][31][32].

\begin{table}[h]
    \centering
    \caption{Compute Specifications}
    
    \begin{tabular}{|l|l|}
        \hline
        \textbf{Item} & \textbf{Specification} \\
        \hline
        CPU & 60TOPS \\
        \hline
        GPU & 50TOPS \\
        \hline
        Bandwidth & 4GBps \\
        \hline
        aiWare Core& 1TMAC/sec  \\
        \hline
        RAM & 5-25MB \\
        \hline
        Memory Bandwidth & 5.3 GBps\\
        \hline
        IO & GigaBit Ethernet\\
        \hline
        IO Bandwidth & 70 Gbps\\
        \hline
    \end{tabular}
\end{table}

We have adopted a \textbf{heterogeneous core architecture} for it has[39]
\begin{itemize}
    \item improved performance
    \item task specific optimization
    \item parallel processing capability
    \item power efficiency
\end{itemize}
Compute intensive areas such as the central chiplet unit and infotainment head unit have been deployed with octa and quad cores whereas the less compute areas have utilised quad and dual cores. Within a compute corelet we have placed a shared L1 cache memory, this enhances the performance because of a higher hit rate than a dedicated cache to every core and it also reduces the latency due to interconnects between compute corelet and memory dielet.

Systems utilising inputs from various cameras perform pixel-level processing operations and hence require graphic processors, video encoders, decoders and image re-sizers. They also include \textbf{Deep Learning Accelerators(DLA)} and \textbf{Performance Vision Accelerators(PVA)}.

The memory dielet specs had been structured to accommodate the  
necessary compute for each chiplet [37]. 3 level cache architecture and flash memory has been employed. An eMMC (embedded multi-media card) interconnection is also provided. External DDR4 memory interfacing is furnished in the IO module.

The IO dielet has been crafted to meet the interconnects and intra-connects requirements of the chiplet package.

\begin{figure}[htbp]
  \centering
  \begin{minipage}[b]{0.5\textwidth}
    \centering
    \includegraphics[width=\textwidth]{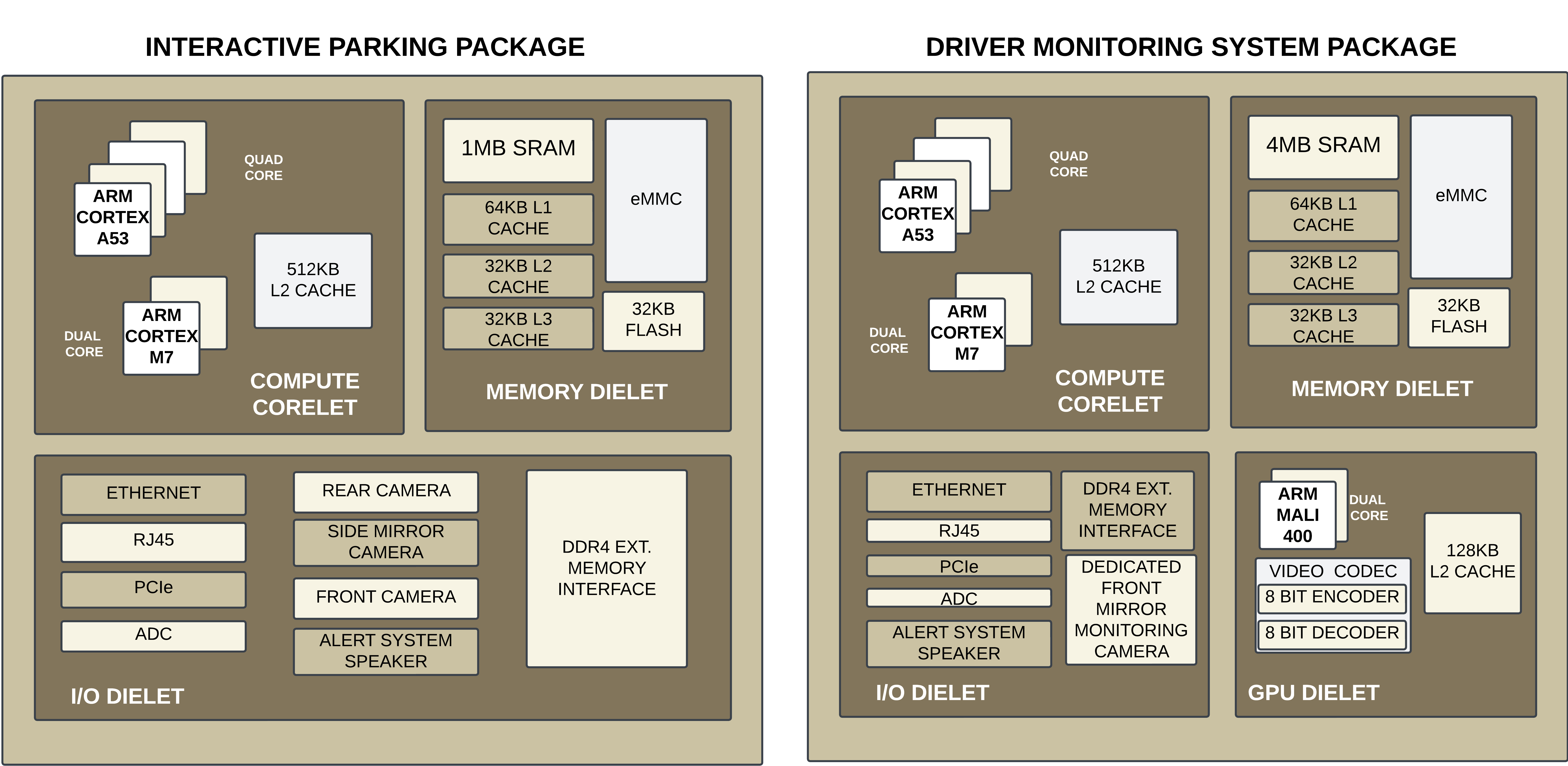}
  \end{minipage}%
  \begin{minipage}[b]{0.5\textwidth}
    \centering
    \includegraphics[width=\textwidth]{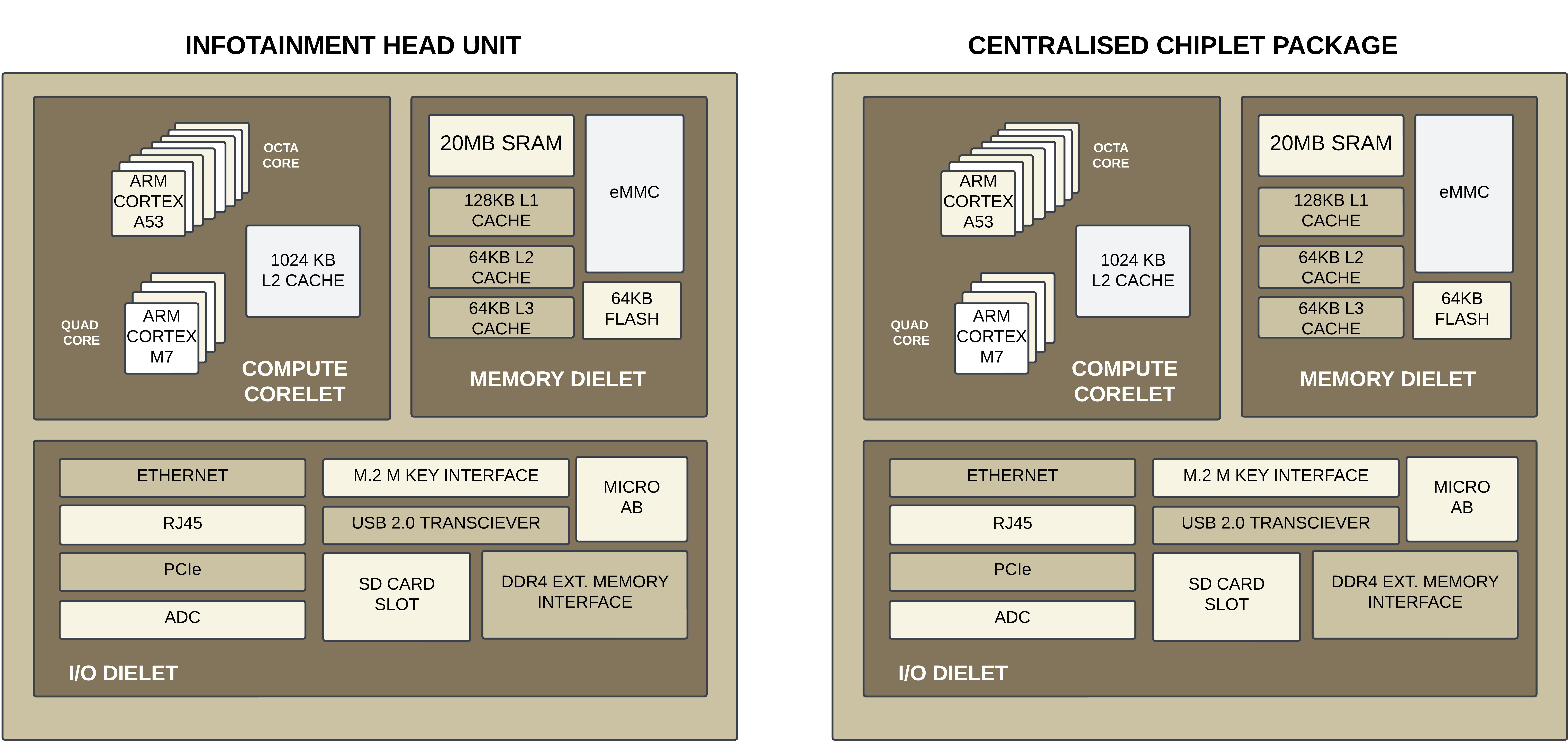}
  \end{minipage}
\end{figure}

\begin{figure}[htbp]
  \centering
  \includegraphics[width=0.5\textwidth]{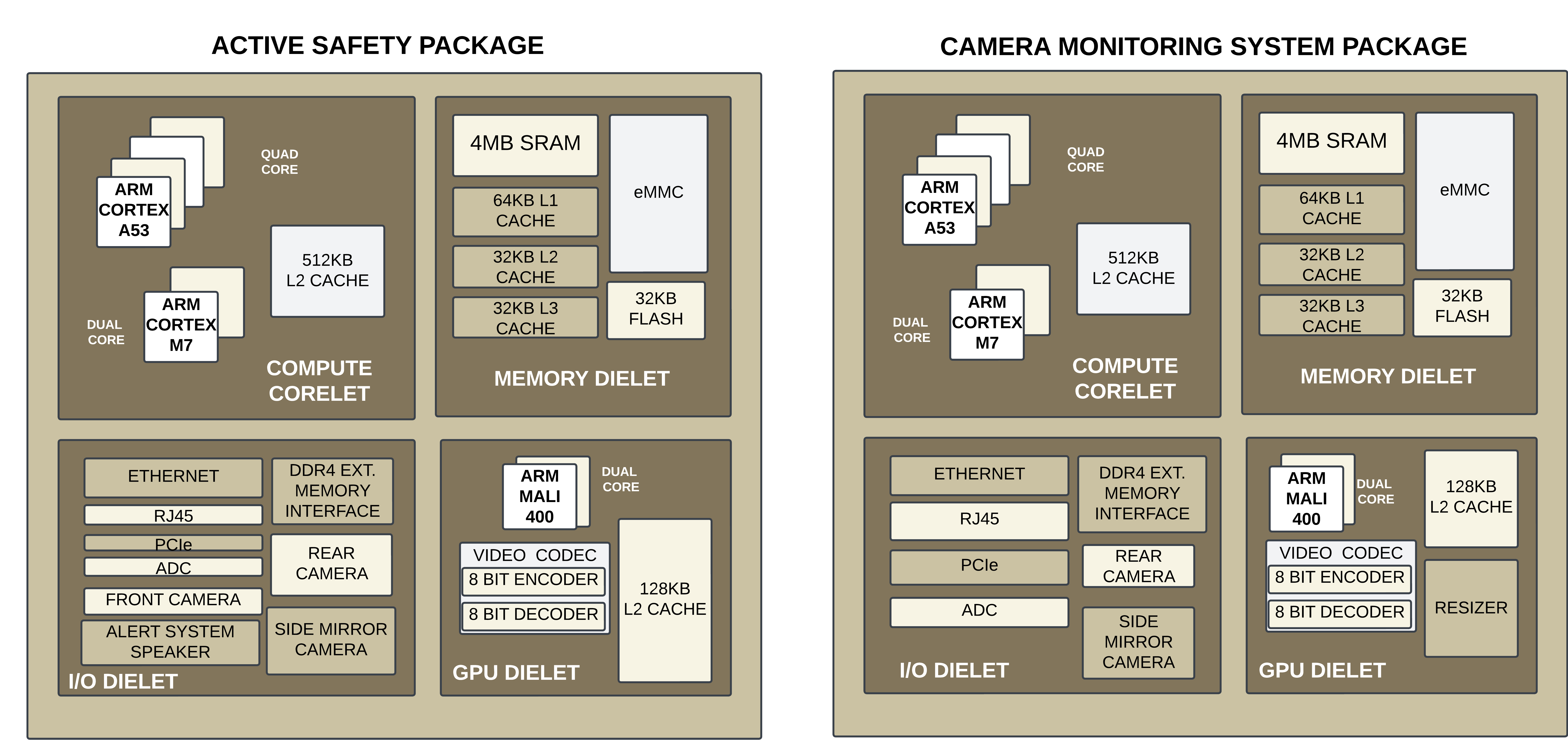}
  \caption{Block level specifications}
\end{figure}

\subsection{Design Considerations \& Constraints}
\begin{itemize}
   \item Compute Corelet and GPU Dielet:

The compute corelet has been designed keeping the \textbf{area as the lower bound} and \textbf{cost the upper bound}. From our cost analysis results, the chiplets are 3.83 times cheaper than an SoC of the same area. Hence, reducing the die sizes reduces the expenses but the optimum limits of area are between 50$mm^2$-150$mm^2$ for microprocessor logic[9]. The area specifications for a die are mentioned below. Assigning 32KB L1 cache per core in the chiplet, we get per core net size of 14.147$mm^2$. This means that about 4 to 10 cores can be placed in a die.

\begin{table}[h]
    \centering
    \caption{Area Specifications}
    \begin{tabular}{|c|c|}
      \hline
      Area per Core & 13.5$mm^2$ \\
      \hline
      Area of 1TMAC aiWare & 25$mm^2$\\
      \hline      
      Area of SRAM 10MB & 69.3$mm^2$ \\
      \hline
      Area of 128KB L1 Cache & 2.589375$mm^2$ \\
      \hline
      Area of 32KB L2 Cache & 4.49$mm^2$ \\
      \hline
      Area of 32KB Flash Memory & 24$mm^2$\\
      \hline
\end{tabular}
\end{table}

The GPU inturn takes in huge amount of compute especially in Neural Network based image and video processing applications, video encoding , decoding , image resizing, perception and  object detection. To aid the the processing power we also use \textbf{aiWare chiplets} that contain \textbf{ML accelerators} as shown in the microarchitecture diagram.

 \item Memory Dielet:
The memory chiplet has been designed keeping the specification requirements as the primary objective. The extensive compute and memory bandwidth in range of GBps requires a large and diverse  memory unit. Shared memory in between the chiplets and dedicated cache memory within the compute and GPU corelet, with caches of appropriate size to maximise hit rate using the random mapping are highlights of the architecture. Again, trading off between cost , power and performance leads us to splitting the memory unit into two identical dielets of size approximately 100$mm^2$. 

\item IO Dielet:
The area specifications of an IO dielet are as follows. Ethernet has been used as the primary chiplet interconnect to give a bandwidth of approximately, 70Gbps. Other pioneering chiplet interconnect and packaging technologies have been discussed extensively further.
\begin{table}[h]
    \centering
    \caption{Area Specifications}
    \begin{tabular}{|c|c|}
      \hline
      NIU (Ethernet) & 7.9317$mm^2$ \\
      \hline
      PCIe & 6.24$mm^2$\\
      \hline      
      ADC & 14$mm^2$ \\
      \hline
      DDR4 & 10.917$mm^2$ \\
      \hline
      MICROAB, USB2.0, M.2 M KEY , SD CARD & 0.92$mm^2$ \\
      \hline
\end{tabular}
\end{table}

\end{itemize}

\subsection{Architecture Simulation}
Now that we have looked into the constraints of designing a chiplet based micro-architecture, We require a precise way of evaluating the throughput-latency performance of the various architectures possible within the given boundaries. For this purposes, We turn to the functional micro-architecture simulator \textbf{gem5} \cite{b45}. The gem5 framework is modified by integrating \textbf{Garnet} - a detailed On-Chip Network Model for Heterogeneous SoCs, thus being able to simulate a variety of chiplet configurations. To simulate synthetic traffic, We use graph algorithms running on the multi-chiplet systems as our benchmark. Our primary objective of this simulation is to evaluate the performance based on the average package latency and throughput of the system.

At its core, gem5 operates on the basis of discrete-event simulation, modeling the execution of instructions and system events with a high degree of fidelity. The simulator's functionality is underpinned by a modular and extensible design, facilitating the representation of diverse microarchitectural features and configurations. The gem5 system assumes a detailed understanding of architectural and microarchitectural components, relying on an event-driven simulation paradigm to capture the intricate interactions among instructions, memory hierarchies, and pipeline stages. Key assumptions within gem5 include the accurate representation of instruction semantics, faithful modeling of memory subsystems, and the provision of flexible configuration parameters for varying microarchitectural parameters. The simulator operates at an abstract level, translating architectural specifications into microarchitectural events, thus enabling researchers to evaluate and optimize the performance of novel processor designs in a controlled and reproducible environment.

We simulate the micro-architecture configurations with the parameters mentioned in the below table.
\begin{table}[h]
    \centering
    \caption{Gem5 Simulation Parameters}
    \begin{tabular}{|c|c|}
      \hline
      CPU type & RISC-V Core \\
      \hline
      L1 private cache size & 32 kB\\
      \hline      
      Workload & BFS with Adjacency matrix \\
      \hline
      Clock Frequency & 2 GHz \\
      \hline
      Area of 32KB L2 Cache & 4.49$mm^2$ \\
      \hline
      Area of 32KB Flash Memory & 24$mm^2$\\
      \hline
\end{tabular}
\end{table}

Based on the amount of compute required and the number of cores that can be fabricated on a die (4-10), we have come up with the following consolidation configurations for the Compute core:
\begin{itemize}
   \item C1: 6 Core + 6 Core
   \item C2: 4 Core + 4 Core  + 4 Core
   \item C3: 8 Core  + 4 Core
\end{itemize}
 Our simulation results for cost, throughput and latency are listed in the table below. We have calculated the \textbf{\textit{Golden Ratio of compute}} as
\begin{equation}
    Golden Ratio = \frac{Throughput}{Latency *  Cost}
\end{equation}

\begin{table}[h]
    \centering
    \caption{Gem5 and Cost Simulation Results}
    \begin{tabular}{|c|c|c|c|c|c|}
      \hline
      \textbf{Config}  & \textbf{Cost} & \textbf{Throughput} & \textbf{Latency} & \textbf{GoldenRatio} &\textbf{Relative} \\
      \hline
      C1  & 129.6854& 1.95E+09 & 30.311 & 4.97E+05 & 1.93\\
      \hline      
      C2 & 177.3822 & 1.97E+09 & 43.234 & 2.58E+05 & 1.00\\
      \hline
      C3  & 136.7064 & 1.92E+09 & 30.763 & 4.58E+05 & 1.78 \\
      \hline
\end{tabular}
\end{table}

From our simulation results it is very clear that the C1 and C3 configurations have approximately same golden ratio of optimization. Since we have adopted a heterogeneous core architecture with ARM CORTEX A53 and ARM CORTEX M7 cores, the 8 Core + 4 Core configuration is better for its ease of \textbf{upgradeability} and \textbf{serviceability}. Hence, we have adopted the same in the micro-architecture.

\subsection{Micro-architecture Illustration}
Based on the abover mentioned considerations, let us therefore visualise the chiplet organization. We have consolidated the compute corelets to have a maximum of 8 cores per chiplet. A shared L1 cache memory placement has been placed inside the compute corelet. On chip SRAM memory of appropriate size and a 3 level cache memory has been fabricated into a memory dielet. All necessary IO port interfaces have been accomodated into the IO dielet as shown.
Chiplet interconnection is achieved through a dedicated \textbf{Network On Chip (NOC) chiplet}. The protocols used for interfacing are UCIe, BOW and HBM protocols.

We present here the micro-architecture diagram with a generalised specification below. This is a differing attribute for each and every chiplet package depending on the spec requirement.
\begin{figure}[htbp]
     \includegraphics[width=\textwidth]{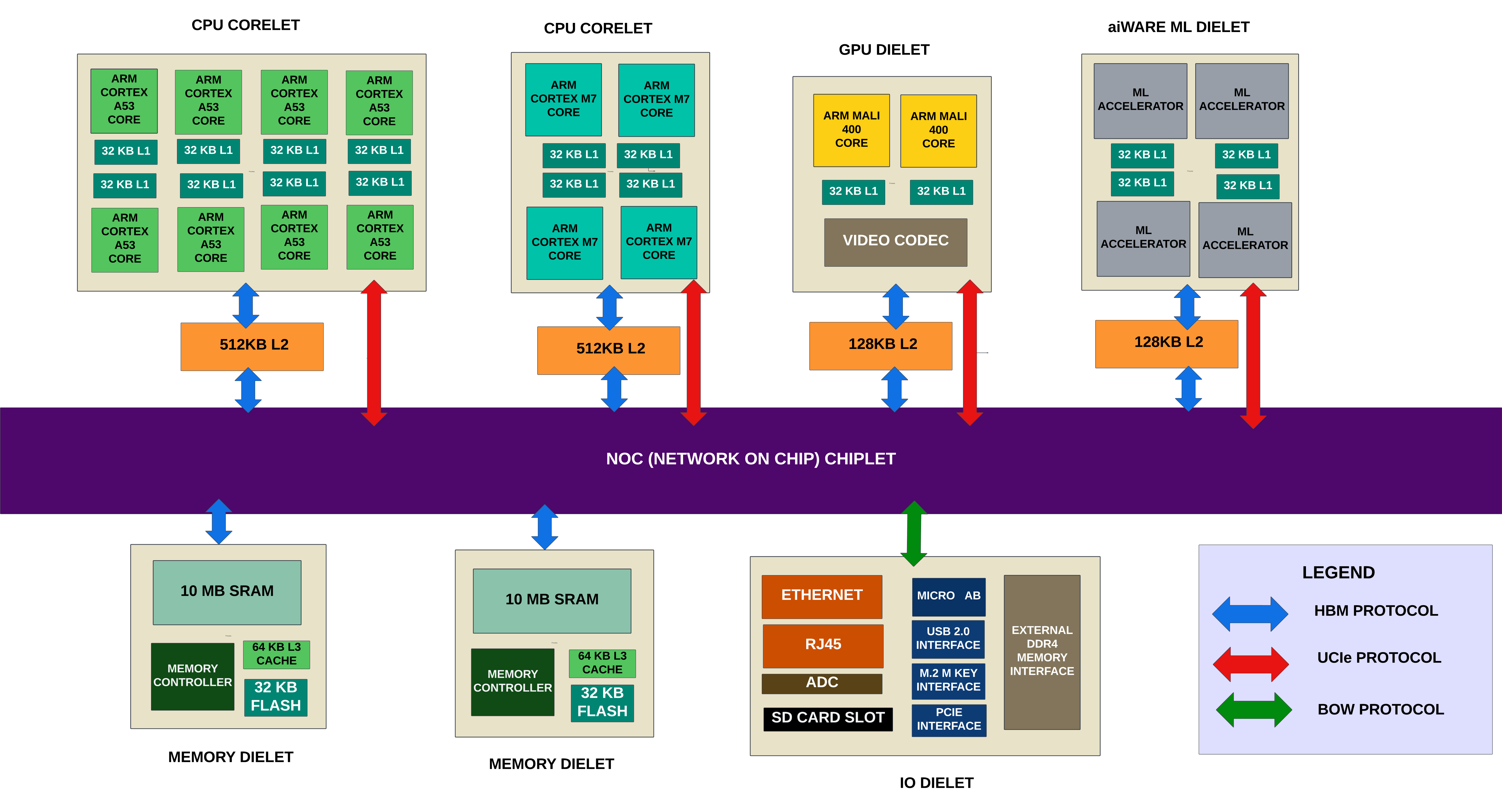} 
     \caption{Micro-architecture diagram} 
\end{figure}

\section{Interconnect Technology}

In a chiplet-based architecture, communication between chiplets is crucial to enable collaboration and data sharing among these specialized, self-contained semiconductor components. It ensures the seamless operation of the overall system, allows for scalability and flexibility, optimizes performance, and supports customized and energy-efficient designs. This inter-chiplet communication is essential for creating complex, high-performance microelectronic systems that can adapt to varying demands and requirements.
Over the years there have been various interconnect technologies and architectures proposed by leading semiconductor conglomerates such as AMD, Arm, ASE, Google, Intel, Meta, Microsoft, Qualcomm, Samsung, and TSMC. We have consolidated the prominent ones and have explained previously in the midterm evaluation. We have summarised the parameters of these interconnect technologies in the table below and have based our arguments for the proposed interconnect on this.

\begin{table}[ht]
\centering
\caption{Interconnect specifications}
\label{table:interconnect1}
\setlength{\tabcolsep}{1pt} 
\begin{tabular}{@{}c c c c c c@{}}
\toprule
\textbf{Interconnect} & \textbf{Bandwith } & \textbf{Latency } & \textbf{Energy Eff. } & \textbf{Channels} & \textbf{Bump density} \\
\textbf{} & \textbf{ (GB/s/mm)} & \textbf{ (ns)} & \textbf{(pJ/b)} & \textbf{} & \textbf{ (micron)} \\
\midrule
UCIe & 28 to 1317 & 2 & 0.25 to 0.5 & 16 to 64 & 25 to 55 \\
AIB & 256 to 1024 & 3.56 & 0.85 & 40 to 160 & 55 \\
BoW & 1280 & 5 & 0.5 to 1 & 16 & 40 \\
QPI & 6.4GT/s & 100 & 10 to 40 & 20 & 400 \\
CXL & 32GT/s & 100 to 200 & 6 & 32 & - \\
Infinity fabric & 5000MT/s & 200 & - & 32 &  - \\
LIPINCON & 1.6Tb/s/$mm^2$ & - & 0.56 & - &  -\\
\bottomrule
\end{tabular}
\end{table}

The escalating computational demands in the automotive industry necessitate the implementation of sophisticated chiplet-based architectures, thereby emphasizing the crucial need for reliable and seamless communication among these components.

The optimal interconnect technology should facilitate high-speed data transfer while simultaneously upholding robustness, security, and efficiency. The following table outlines the specifications of existing chiplet interconnect technologies. Our proposal advocates for the utilization of distinct interconnect technologies tailored to specific component sets within the chiplet.

UCIe, a novel standard developed by Intel, stands out as the most promising candidate to emerge as the universal standard for chiplet interconnects, given its endorsement by major chip vendors. We recommend UCIe due to its interoperability with leading chip vendors and its compelling specifications. With latency below 2ns and the best energy efficiency (0.25 pJ/b) among the discussed interconnects, UCIe proves ideal for integrating the digital aspects of the chiplet, such as processors and accelerators, where high data rates are essential.

For digital-memory and memory-memory communication, we propose the use of HBM2 and HBM3. HBM technology, characterized by vertically stacked memory chips, enables smaller form factors, ultra-wide communication lanes, a wider memory bus, and higher bandwidth. Given that memory fetch and store operations incur significant latencies, HBM emerges as an optimal solution for memory communication.

In the realm of analog-digital communication, we recommend the adoption of BoW. BoW offers a less complex packaging design without the need for interposers. As a parallel bus protocol, BoW is energy-efficient and features denser wire packing. While it competes favorably with UCIe in terms of efficiency and bandwidth, UCIe boasts superior throughput and greater flexibility in packaging. Consequently, we propose the use of BoW for peripherals and analog-digital interfaces.

\begin{table}[ht]
\centering
\caption{Summary}
\label{table:interconnect2}
\setlength{\tabcolsep}{1pt} 
\begin{tabular}{@{}c c @{}}
\toprule
\textbf{Interconnect} & \textbf{Application }\\
\midrule
UCIe & Digital interfaces(processors, cores , accelerators etc) \\
BoW & analog-digital interfaces(ADC, sensors etc) \\
HBM & Memory interfaces\\
\bottomrule
\end{tabular}
\end{table}

\section{Interconnect Electrical Layer Analysis}
\subsection{Physical Layer Analysis}
With the choice of interconnect established above, it is critical to derive the physical layer parameters and limits of the interconnect technology to match the required clock frequency and data rate. These limits would also set the baseline for the system-level packaging of chiplet. In this regard, the physical layer implementation of the interconnect is characterized as follows:

\begin{itemize}
    \item The interconnect is packaged using 2.5D technology with a Silicon - Silicon-dioxide interposer that is $100\mu m$ thick with a relative permittivity $\epsilon_r = 11.68$.
    \item The interconnect trace is made of Copper (conductivity $\sigma_{Cu}=5.98\times 10^7 S/m$. The trace is $50\mu m$ wide and $20\mu m$ thick. It is modelled as a stripline trace with a ground plane thickness of $50\mu m$.
    \item The system is clocked at $2GHz$ frequency and the interconnects have to be operational at this frequency.
\end{itemize}

The minimum required $3dB$ bandwidth of the interconnect is thus $2GHz$, but considering noises and clock PLL errors, a safety factor of $1.5$ is chosen.

\[SF = 1.5\]
\[3dB \: BW \: (Target) = SF \times Clock \: Frequency\]
\[3dB \: BW \: (Target) = 1.5 \times 2GHz = 3 \times 10^9 Hz\]

The bandwidth is ultimately limited by the rise time ($T_r$) and fall time ($T_f$) of the signal on the interconnect line. For symmetric operation, $T_r = T_f$. The 3dB bandwidth of a digital channel is given by \cite{b27}

\[3dB \: BW = \frac{0.35}{T_r}\]

The rise time $T_r$ is essentially the time taken for the rising edge to go from 10\% to 90\% of VDD. This is given by:

\[T_r = R_{tr}C_{tr} \times ln\left(\frac{9}{1}\right)\]

where, $R_{tr}$ and $C_{tr}$ are the equivalent resistance and capacitance of the trace. These parameters are dependent on the trace dimensions and substrate (interposer) material. The objective is the find the maximum trace length that allows us to meet our bandwidth target. The models used for per length resistance and capacitance of a stripline trace are as mentioned in \cite{b28}. From these models, it was calculated that:

\[C_{tr} = 389 \times l_{tr} \: pF\]
\[R_{tr}(DC) = 16.72 \times l_{tr} \: \Omega\]
\[R_{tr}(AC) = 85.64 \times l_{tr} \: \Omega\]
\[R_{tr} = R_{tr}(DC)+R_{tr}(AC) = 102.36 \times l_{tr} \: \Omega\]

where, $l_{tr}$ is the length of the trace. Using the above numbers, we can derive the 3dB bandwidth of the interconnect channel as a function of $l_{tr}$. The log plot of this 3dB BW and the target 3dB BW is given below:

\begin{figure}[htbp]
    \centering
    \includegraphics[width=0.5\textwidth]{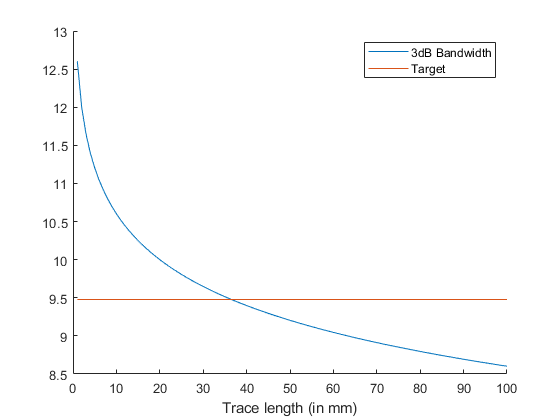} 
    \caption{Log Plot of Bandwidth against Trace Length} 
\end{figure}

From the plot, it can be inferred that the intersection point ($36mm$), is the maximum possible distance between 2 interconnected chiplets on the interposer. 

\begin{multicols}{2}

\begin{verbatim}
u0 = 1.2566e-6
e0 = 8.8542e-12
v0 = 1/(u0*e0)^0.5

%%% Trace Capacitance
Er = 11.68 % Silicon Interposer
interposerWidths = 100e-6
width = 50e-6
capPerLength = Er*(width/interposerWidths+
0.441)/(30*pi*v0)

%%% Trace Resistance
clkFreq = 2e9
thickness = 20e-6
gndThickness = 50e-6
sigma = 5.98e7
skinDepth = (pi*clkFreq*u0*sigma)^-0.5
resDCPerLength = 1/sigma*(1/(width*thickness)
+1/(2*gndThickness))
resACPerLength = 1/sigma*(1/(skinDepth*
(2*thickness-4*skinDepth+2*width))+1/(2*sigma))
resPerLength = resDCPerLength + resACPerLength

%%% Bandwidth
BWSafetyFactor = 1.5
threeDBBWTarget = log10(BWSafetyFactor*clkFreq)
threeDBBWTarget = linspace(threeDBBWTarget, 
threeDBBWTarget, 100)
length = linspace(1e-3, 100e-3, 100)
riseTime = resPerLength*capPerLength*length
.^2*log(9)
threeDBBW = log10(0.35./riseTime)
\end{verbatim}

\end{multicols}

\subsection{Latency \& Bandwidth Analysis}
We simulate for various configurations of bandwidth of the serial protocol using the \textbf{gem5} simulator we explained before.

\begin{figure}[htbp]
  \centering
  \begin{minipage}[b]{0.5\textwidth}
    \centering
    \includegraphics[width=\textwidth]{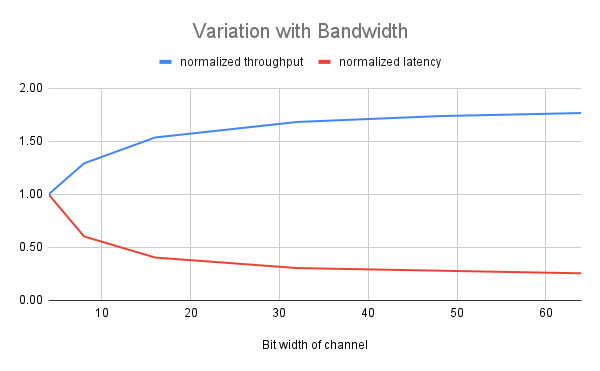}
    \caption{Variation with Communication bandwidth}
  \end{minipage}%
  \begin{minipage}[b]{0.5\textwidth}
    \centering
    \includegraphics[width=\textwidth]{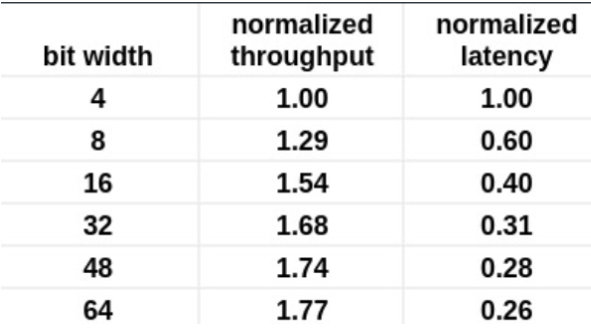}
    \caption{Variation with Communication bandwidth}
  \end{minipage}
\end{figure}

The results are seen in the figure below. Increasing the bandwidth by twice doesn't really yield a two fold increase in throughput or two fold decrease in latency. That relation depends on the implementation and the specification followed.

Below are the results obtained for a general serial communication protocol, only from a functional simulation perspective. The relationship is not linear given it is for a general protocol. UCIe and BoW are such that their specifications yield good throughput, less latency at high bandwidth. The literature numbers and test simulations back this up. 

\begin{figure}[htbp]
  \centering
  \begin{minipage}[b]{0.5\textwidth}
    \centering
    \includegraphics[width=\textwidth]{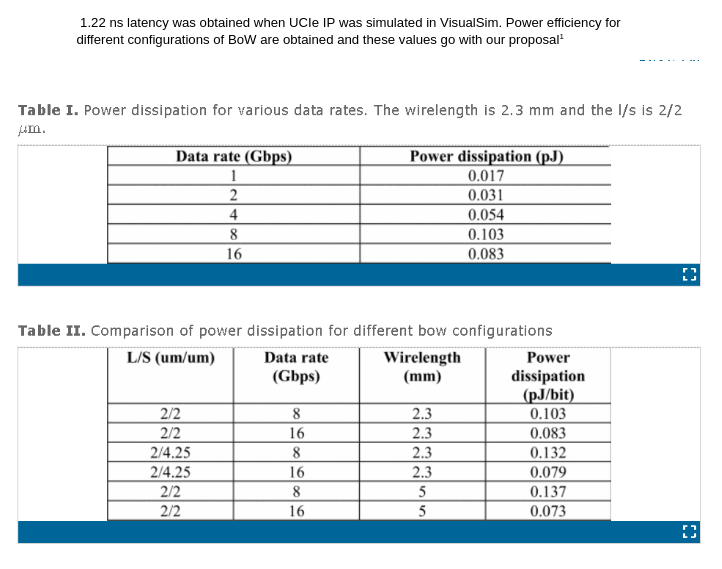}
    \caption{Variation with Communication bandwidth}
  \end{minipage}%
  \begin{minipage}[b]{0.5\textwidth}
    \centering
    \includegraphics[width=\textwidth]{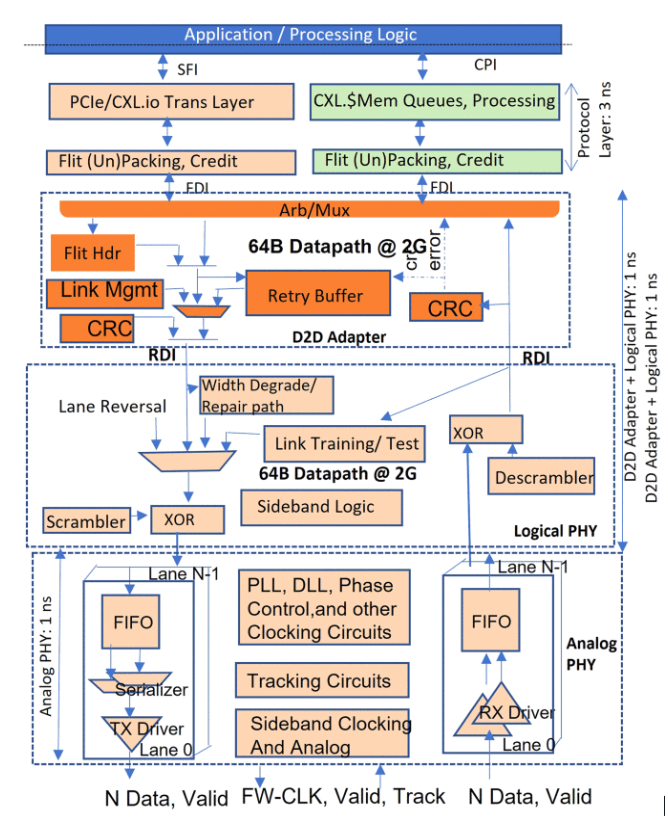}
    \caption{Variation with Communication bandwidth}
  \end{minipage}
\end{figure}

People have achieved a bandwidth of 32Gb/s/wire and 0.44pJ/bit energy efficiency by implementing a die-to-die (D2D) chiplet compatible with the UCle specification using 2.5D packaging technology. This is explained in the paper “A 4nm 32Gb/s 8Tb/s/mm Die-to-Die Chiplet Using NRZ Single-Ended Transceiver With Equalization Schemes And Training Techniques”. Data centre and AI group form Intel have implemented UCIe and obtained the below latency values.

\section{Security \& Reliability}
Any chiplet being designed and manufactured for automotive purposes has to comply with the ISO 21434: 
\begin{itemize}
    \item Security Requirements:

    The standard defines a set of security requirements that semiconductor manufacturers must meet. These requirements cover aspects such as secure design, development, manufacturing, and supply chain management.
    The requirements aim to minimize vulnerabilities and ensure the integrity of semiconductor components used in automotive systems.
    Some specific requirements include:
    \begin{itemize}
        \item Implementing secure boot and firmware update mechanisms.
        \item Utilizing encryption and authentication protocols for secure communication.
        \item Employing hardware security modules (HSMs) for key management and cryptographic operations.
        \item Conducting vulnerability assessments and penetration testing to identify and address potential security weaknesses.
    \end{itemize}

\item Threat Assessment and Risk Management:
\begin{itemize}
    \item The standard requires manufacturers to conduct thorough threat assessments to identify potential vulnerabilities and cyberattacks their semiconductors may face.
    \item This includes analyzing the attack surface, potential threats, and the impact of successful attacks on vehicle safety and security.
    \item Based on the threat assessment, manufacturers must implement risk management measures to mitigate identified risks. This may involve prioritizing vulnerabilities, implementing additional security controls, and developing incident response plans.
\end{itemize}

\item Security Assurance:
\begin{itemize}
    \item ISO 21434 defines a set of security assurance activities manufacturers must perform to demonstrate compliance with the standard.
    \item These activities may include code reviews, penetration testing, security audits, and vulnerability assessments.
    \item Third-party certification bodies can also be involved to verify the implementation of security controls and ensure compliance with the standard.
\end{itemize}

\item Supply Chain Security:
\begin{itemize}
    \item The standard recognizes the importance of a secure supply chain in ensuring the overall security of automotive systems.
    \item It requires manufacturers to implement robust security practices throughout their supply chain, including working with trusted suppliers and implementing security controls at all stages of production.
\end{itemize}

\item Continuous Monitoring and Improvement:
\begin{itemize}
    \item ISO 21434 emphasizes the need to monitor and improve cybersecurity practices continuously.
    \item Manufacturers must monitor their systems for vulnerabilities and security threats and continuously update their security controls to adapt to evolving threats.
    \item This requires a proactive approach to cybersecurity and a commitment to continuous improvement.
\end{itemize}
    
\end{itemize}
The security vulnerabilities chiplets offers are as follows:
\begin{itemize}
    \item Increased Attack Surface:

    More interfaces and communication paths compared to monolithic chips create a larger attack surface.
    Attackers can exploit vulnerabilities in these interfaces to gain unauthorized access, manipulate data, or launch denial-of-service attacks.

\item Hardware Trojans:

    Malicious circuits can be embedded in individual chiplets during the manufacturing process, compromising the entire system.
    These Trojans are difficult to detect and remove and can be used for data theft, disrupting functionality, or launching physical attacks.

\item Supply Chain Risk:

    Reliance on multiple vendors for chiplets increases the risk of vulnerabilities introduced during design, manufacturing, or assembly.
    Malicious actors can infiltrate the supply chain to tamper with chiplets, introducing vulnerabilities or backdoors.

\item Software Vulnerabilities:

    Each chiplet may run its own software, potentially introducing vulnerabilities.
    These vulnerabilities can be exploited for unauthorized access or compromising the entire system.

\item Insecure Inter-Chiplet Communication:

    Communication between chiplets often occurs through interposers or other high-speed interfaces.
    If not secured, these interfaces are vulnerable to eavesdropping, data manipulation, and denial-of-service attacks.

\item Man-in-the-Middle Attacks:

    Attackers can intercept communication between chiplets to steal data or inject malicious code.
    This is particularly relevant for chiplets not physically protected within a secure enclave.

\item Physical Attacks:

    Chiplets are susceptible to physical attacks like side-channel analysis and fault injection.
    These attacks can extract sensitive information or disrupt chiplet functionality.

\item Counterfeit Chiplets:

    Malicious actors can manufacture and sell counterfeit chiplets containing vulnerabilities or backdoors.
    This compromises the entire system's security if not detected and removed.

\end{itemize}

We propose to divide the security in 3 stages:
\begin{itemize}
    \item \textbf{Chiplet Level} \\
    \item Secure Boot and Root of Trust (RoT):

    Secure Boot: Ensures that only authorized firmware is loaded upon chiplet startup, preventing unauthorized code execution and malicious manipulation.\\
    Root of Trust: Acts as a central trust anchor within the chiplet, providing secure storage for cryptographic keys and facilitating secure communication with other components.
    \item \textbf{Hardware Security Modules (HSMs)}:

    Dedicated hardware modules designed to perform cryptographic operations efficiently and securely. That is by implementing the CHSM ( Chiplet Hardware Security Module). The module should be responsible for the following:
    \begin{itemize}
        \item Blocking secure communication between IPs in a chiplet.
        \item Stopping the propagation of any malicious signal outside a chiplet.
        \item Masking the injected faults by designing hardware patches. 
        \item Blocking the fault injection means detecting the chiplet level tampering.
    \end{itemize}
    \item \textbf{Communication }
    We have devised a thermal covert channel to communicate between processors for sensitive data like secret keys.
    The size of data that has to be sent from one processor to another processor is defined before communicating. Each processor is connected to a ring oscillator design. When the bit that has to be sent is high (logic 1). A trigger signal will be sent through which the ring oscillator undergoes continuous switching. Due to a lot of bit flips during a very short amount of time, the temperature of the device increases. The bit flips happen for a few ms[Fig ]. We have emulated this feature on a ZCU-102 FPGA and achieved a bit of transmission rate of \textbf{2bps}. We ran an FFT algorithm to detect the peaks with a variable window according to our needs. This mode of communication can be used to communicate secret keys or sensitive data to prevent man-in-the-middle attacks. 

\begin{figure}[h]
  \centering
  \begin{minipage}[b]{0.5\textwidth}
    \centering
    \caption{ Temperature measurements recorded for a thermal covert channel to send data}
    \includegraphics[width=\textwidth]{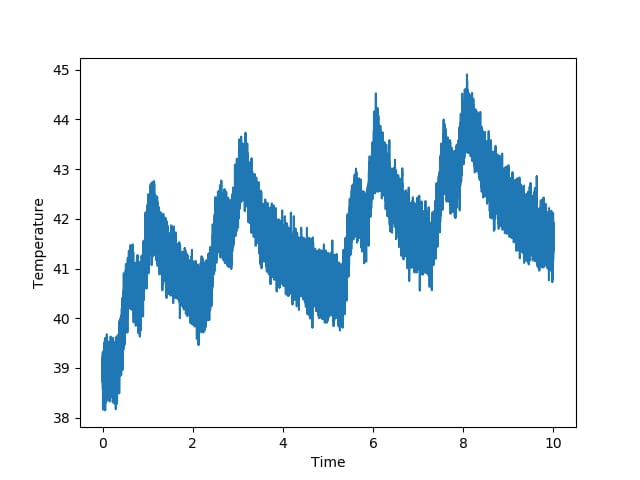}
  \end{minipage}%
  \begin{minipage}[b]{0.5\textwidth}
    \centering
    \caption{FFT algorithm to identify the bits being sent}
    \includegraphics[width=\textwidth]{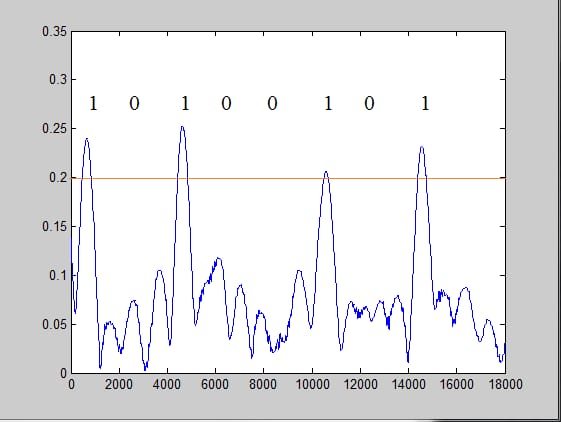}
  \end{minipage}
\end{figure}

\item \textbf{SiP level}
\begin{itemize}
    \item Access Control Policies: Access control policies specify
how one chiplet can access an asset during different execution
points for the SiP. These are:
\begin{itemize}
    \item An unauthorized chiplet cannot access memory in the
protected address range.
\item An unauthorized chiplet cannot write out data to a restricted memory region (information leakage).
\end{itemize} 
\item Information Flow Policies: Information Flow Policies
restrict leakage or modification of information related to secure
assets. Examples of such policies are:
\begin{itemize}
    \item An unauthorized chiplet cannot access data intended for
other chiplets during transit.
\item An unauthorized chiplet cannot modify data intended for
other chiplets.
\item Chiplet A cannot pose as chiplet B to receive data
intended for chiplet B.
\item Data intended for a chiplet cannot be blocked by an
unauthorized chiplet.
\end{itemize}

\item Liveness Policies: Liveness policies ensure that SiP can
execute normal tasks without interruption. Examples include:
\begin{itemize}
    \item A chiplet cannot flood communication fabric with messages to disrupt normal behavior (DDoS).
\item During operation, the number of messages sent by an
untrusted chiplet should not exceed the threshold of the
maximum limit.
\item The limit on the number of packets generated by an
untrusted chiplet can only be assigned and updated at
secure boot time.
\end{itemize}
\end{itemize}

\end{itemize}
\section{Interconnect Packaging}

Interconnect packaging technologies play a critical role in the realm of semiconductor design and system integration. One prominent strategy within this domain is the application of heterogeneous integration using chiplets. This methodology involves the amalgamation of diverse technologies, including digital processors, accelerators, analog components, and memory, with the aim of constructing a system that is not only efficient but also highly flexible.\cite{b2}

Addressing the multifaceted requirements inherent in heterogeneous integration mandates the utilization of distinct chiplet interface protocols. These protocols serve as the linchpin for integrating chiplets endowed with varying functionalities. For instance, the employment of protocols like AXI facilitates the integration of individual cores into cohesive clusters, while low-level protocols come into play for the integration of analog components with processors. Memory integration, on the other hand, leverages protocols such as HBM and LPDDRx to establish seamless connections with processors.

Within the heterogeneous integration framework, each technology fulfills specific roles in the overall system functionality. Digital processors take on the responsibility of executing instructions and performing calculations, accelerators serve as specialized components enhancing the performance of specific tasks, analog components manage signals and furnish interfaces for external devices, and memory assumes the crucial role of storing data and instructions for timely access by the system.\cite{b2}

The conceptualization of chiplets and their integration through the heterogeneous approach aligns harmoniously with the industry's overarching objective to extend Moore's Law. By adopting this strategy, the disaggregation of System-on-Chip (SoC) functionality becomes feasible. This, in turn, leads to the optimization of chiplet performance, the improvement of power efficiency, and the augmentation of flexibility through standardized interfaces. The seamless integration of chiplets, encompassing digital processors, accelerators, analog components, and memory, facilitates configurability and adaptability, contributing to the advancement of semiconductor design and system integration.

Connecting these chiplets poses a significant challenge, leading to a surge of interest and development in the chiplet domain. Various interconnect technologies exist, with Peripheral Component Interconnect Express (PCIe) being a notable example. PCIe, a high-speed serial computer expansion bus standard, serves chip-to-chip (C2C) or board-to-board (B2B) communications. However, for die-to-die (D2D) interconnect, alternatives like Compute Express Link (CXL), Bunch of Wires (BoW), and Universal Chiplet Interconnect Express (UCIe) have gained prominence.

A groundbreaking player in this field is Eliyan, a company that has pioneered NuLink, a physical layer for D2D connections. NuLink is a superset of the UCIe protocol layer and the BoW physical layer, offering intriguing advantages over existing solutions.

\subsection{NuLink: A Novel Interconnect Packaging Technology}

Eliyan's NuLink technology addresses key limitations associated with chiplet-based systems, particularly those using silicon interposers. While silicon interposers provide high trace density, enabling high bandwidth at low power, they come with drawbacks such as limited package size, high cost, and thermal challenges.

NuLink takes a different approach by allowing chiplets to be mounted directly onto a standard organic substrate, \textbf{eliminating the need for silicon interposers}. This results in numerous advantages, including \textbf{larger and more complex systems in a package}, \textbf{lower packaging cost}, \textbf{shorter production cycles}, \textbf{high test coverage}, and \textbf{improved thermal performance}.\cite{b3}\cite{b4}\cite{b5}

\subsubsection{Performance Metrics and Implications for AI ASICs}

Eliyan Technology has introduced NuLink PHY, a chiplet interconnect technology, which, based on a superset of industry standards \textbf{UCIe} and \textbf{BoW}, provides similar bandwidth, power, and latency to interconnects on a silicon-based interposer but on standard organic substrates. This technology allows for increased memory capacity, potentially doubling the amount of memory for AI applications. NuLink PHY enables more HBM memory per ASIC, with the possibility of increasing the number of HBMs by a factor of two. Eliyan's NuLinkX extends the reach of NuLink by 10x to at least 20cm, supporting chip-to-chip external routing over a Printed Circuit Board (PCB), enhancing design flexibility for high-performance systems. The innovations by Eliyan aim to address the challenges faced by high-performance computing (HPC), particularly in the realm of AI processing, by providing more efficient interconnect solutions and increasing memory capacity for memory-dense applications.

\begin{figure}[htbp]
  \centering
  \begin{minipage}[b]{0.5\textwidth}
    \centering
    \includegraphics[width=\textwidth]{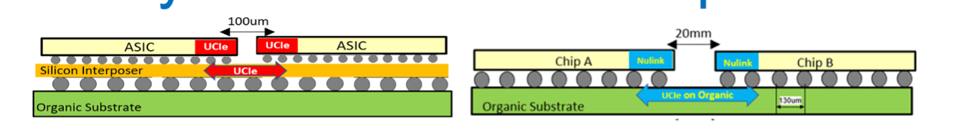}
    \caption{Silicon Interposer versus NuLink's packaging Technology}
  \end{minipage}%
  \begin{minipage}[b]{0.5\textwidth}
    \centering
    \includegraphics[width=\textwidth]{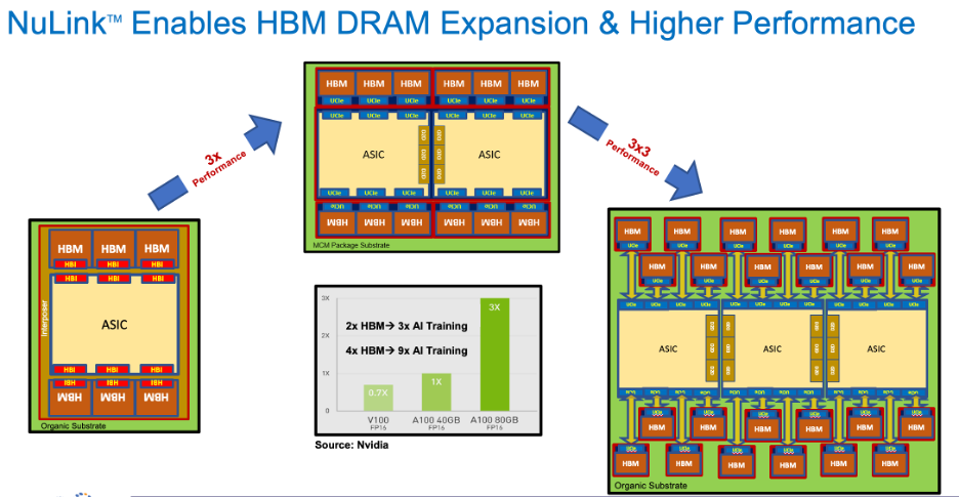}
    \caption{Eliyan NuLink enhances multi-chip designs, enabling increased HBM memory}
  \end{minipage}
\end{figure}

\subsubsection{Economic Implications and Industry Recognition}

Eliyan's chiplet approach offers economic benefits for IC design, making multi-die approaches more attractive for chip suppliers in the automotive sector. This aligns with the industry's need to optimize power and bandwidth, especially for applications like accelerated server computing in data centres.

Industry analysts, exemplified by John Lorenz, Senior Analyst at Yole Intelligence, acknowledge the economic advantages of adopting a chiplet approach. Eliyan's NuLink chiplet interconnect technology is positioned to address the power and bandwidth optimization needs of chip suppliers in automotive computing applications.

\subsubsection{Advancements in System-in-Package (SiP) Solutions}

The ability to implement chiplet-based systems in standard organic packages empowers the creation of larger System-in-Package (SiP) solutions for the automotive sector. This facilitates higher performance per power, considerably lower cost, and higher yield.

NuLink's compatibility with multiple foundry and node technologies, based on early customer interest and demand, highlights its flexibility in meeting diverse automotive application requirements.

\subsubsection{Path Forward and Industry Impact}

NuLink's foundation of the Bunch of Wires (BoW) standard, adopted by the Open Compute Project, and compatibility with the UCIe standardization efforts emphasize its collaborative approach within the industry.

Ongoing efforts to create an efficient universal die-to-die interconnect optimized for memory traffic underscore NuLink's commitment to future-focused developments. This aims to accelerate the adoption of memory chiplets in automotive computing.

Up to this point, our discussion has centered on interconnect technology that excludes the use of an interposer. Redirecting our focus to designs incorporating an interposer, we highlight a groundbreaking technology employed by ASICLAND, where they have implemented the RDL interposer in their innovative design.

\subsection{RDL Interposer}

\subsubsection{Introduction}

The semiconductor industry's pursuit of enhanced performance, reduced size, and lower manufacturing costs has led to the exploration of advanced packaging technologies. The 2.5D RDL interposer emerges as a key player in fan-out wafer level packaging (FOWLP) development, catering to the demand for cost-attractive solutions for heterogeneous chip integration.

\subsubsection{Key Features of 2.5D RDL Interposer}

The proposed 2.5D RDL interposer technology introduces a fine-pitch RDL interposer (\textgreater 560 mm\textsuperscript{2}) capable of accommodating one high-bandwidth memory (HBM) and two ASICs. The design aims to achieve a (through-silicon via) TSV-less and cost-effective package. The fine-pitch RDL interposer enhances signal integrity and bump joint reliability, allowing for the integration of multiple chips with higher I/O counts.

 \begin{figure}[htbp]
     \centering
     \includegraphics[width=0.7\textwidth]{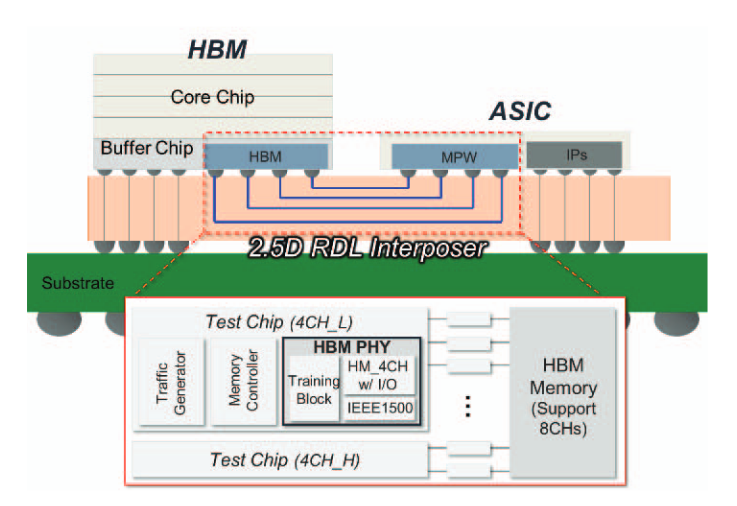} 
     \caption{Package architecture of 2.5D RDL interposer} 
 \end{figure}

\subsubsection{Performance Evaluation}

The fine-pitch 2.5D RDL interposer package demonstrates robust performance, achieving up to 3.2 Gbps/pin operation with the HBM. The reliability tests, including TC1000hr, b-HAST 264hr, u-HAST 264hr, and HTS1000hr, reveal excellent reliability without any failures. The proposed technology positions itself as a promising solution for cost-effective and large-size 2.5D packaging in high-performance computing (HPC) applications.\cite{b7}

\begin{table}[h]
    \centering
    \caption{Specifications of the Package Structure}
    
    \begin{tabular}{|l|l|}
        \hline
        \textbf{Item} & \textbf{Specification} \\
        \hline
        Device & HBM, Logic \\
        \hline
        Number of Dies & 3 (1 HBM, 2 Logics) \\
        \hline
        Minimum L/S of RDL & 2/2 $\mu$m \\
        \hline
        Size of Package & 42.5 x 42.5 mm\textsuperscript{2} \\
        \hline
        Interposer & \textgreater\ 560 mm\textsuperscript{2} \\
        \hline
        Number of Joints (Chip) & \textgreater\ 9,000 \\
        \hline
        Number of Joints (Interposer) & \textgreater\ 20,000 \\
        \hline
    \end{tabular}
\end{table}

\vspace{0.5cm}
\subsubsection{Comparison with Silicon Interposer}

In the realm of interconnect packaging, silicon interposers have been prominent but come with limitations such as high cost, thermal challenges, and limited package size. The 2.5D RDL interposer, in contrast, offers advantages like high productivity, lower cost, and good reliability. Its intrinsic properties, including low latency in chip-to-chip communication, low thermal resistance, and design flexibility, make it a compelling choice for advanced packaging in HPC applications.

\subsubsection{Experimental Insights}

The experiments conducted to validate the 2.5D RDL interposer's functionality include the fabrication of a package structure with one HBM and two logic chips. The package specifications, including size, minimum L/S of RDL, and the number of joints, contribute to its effectiveness. The proposed fabrication process involves RDL fabrication, multi-chip bonding, encapsulation, chip exposure, solder ball attachment, and interposer assembly on the substrate.

\subsubsection{Functionality and Reliability Evaluation}

Signal integrity simulation is a critical aspect of evaluating the HBM-PHY functionality. The simulations showcase the optimized signal routing arrangements, resulting in increased eye-opening values after optimization. The fine-pitch 2.5D RDL interposer demonstrates read and write VWM operations at 3.2 Gbps/pin, validating its functionality.

Reliability evaluation, including TC, b-HAST, u-HAST, and HTS tests, confirms the robustness of the 2.5D RDL interposer. The stress distribution analysis during TC cycles reveals a unique stress concentration at the edge of the chip, showcasing the structural integrity of the proposed system.

The 2.5D RDL interposer technology emerges as a promising solution for advanced packaging in HPC applications. Its advantages over silicon interposers, coupled with strong experimental and simulation results, position it as a key enabler for the new era of heterogeneous chip integration. The economic implications, reliability, and performance metrics underscore the significance of the 2.5D RDL interposer in the evolving landscape of interconnect packaging.

\subsubsection{ASICLAND's Advancements in Chip Packaging Technology}

Chip design house ASICLAND is actively working on developing a new packaging technology aimed at enhancing the cost efficiency of TSMC's chip-on-wafer-on-substrate (CoWoS) technology.

At a local conference hosted by TheElec in Seoul, ASICLAND manager Kang Sung-mo highlighted the potential for performance and power consumption improvements in CoWoS compared to integrated fan-out and organic substrate packages.

CoWoS, known for its utilization of silicon interposer and through-silicon via (TSV), provides flexibility in controlling the interposer size. Kang emphasized ASICLAND's focus on improving the cost of the silicon interposer through the development of a new packaging solution.

\subsubsection{ASICLAND's RDL Interposer Package}

ASICLAND's innovative approach involves the use of a Redistribution Layer (RDL) interposer, addressing cost concerns associated with traditional silicon interposers. This new package configuration places HBM (High Bandwidth Memory) and SoC (System-on-Chip) on top of the RDL interposer, with a silicon bridge facilitating the connection between the dies. Additionally, ASICLAND has incorporated a heat spreader into the design for improved thermal management.

\subsection{Apple's Approach with UltraFusion}
Let's examine the advancements that Apple has made to their chips thus far.\cite{b8}

Apple revolutionised the silicon industry with the introduction of their custom silicon used in their computers and 
The M1 Ultra processor utilizes a custom-built packaging architecture that connects the dies of two M1 Max processors. This connection is achieved using TSMC's CoWoS-S technology, a silicon interposer that links the dies of the two M1 Max chips. It is important to note that CoWoS-S is a more expensive technology compared to TSMC's InFO\_LSI, which uses localized silicon bridges instead of full silicon interposers. However, InFO\_LSI was not available in time for the M1 Ultra, leading Apple to choose the proven but pricier CoWoS-S solution.

There is speculation among experts that Apple might adopt InFO\_LSI in the next generation of the M1 processor. 

The M1 Ultra's substrate is supplied by Unimicron and features ABF RDL. The bump pitch of the M1 Ultra is 25µm. It is categorized as a System-on-Chip-in-Package (SoCiP) as it combines two M1 Max chips.

\begin{figure}[htbp]
  \centering
  \begin{minipage}[b]{0.5\textwidth}
    \centering
    \includegraphics[width=\textwidth]{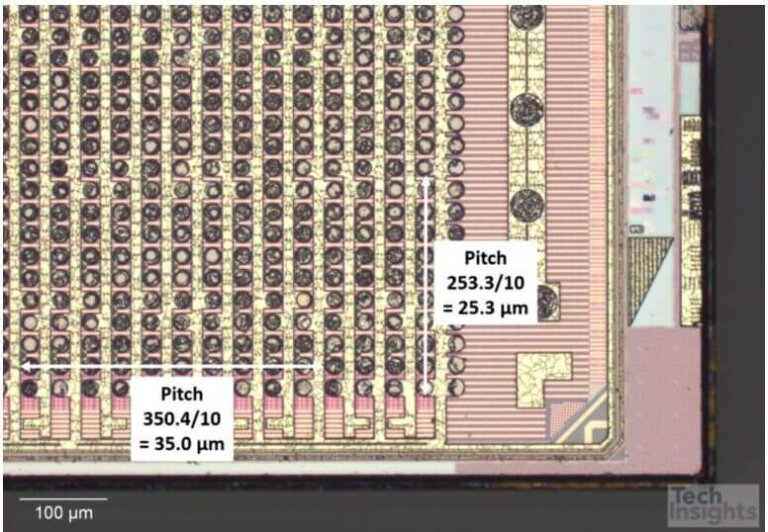}
    \caption{TearDown of Apple M1 Ultra chip}
  \end{minipage}%
  \begin{minipage}[b]{0.5\textwidth}
    \centering
    \includegraphics[width=\textwidth]{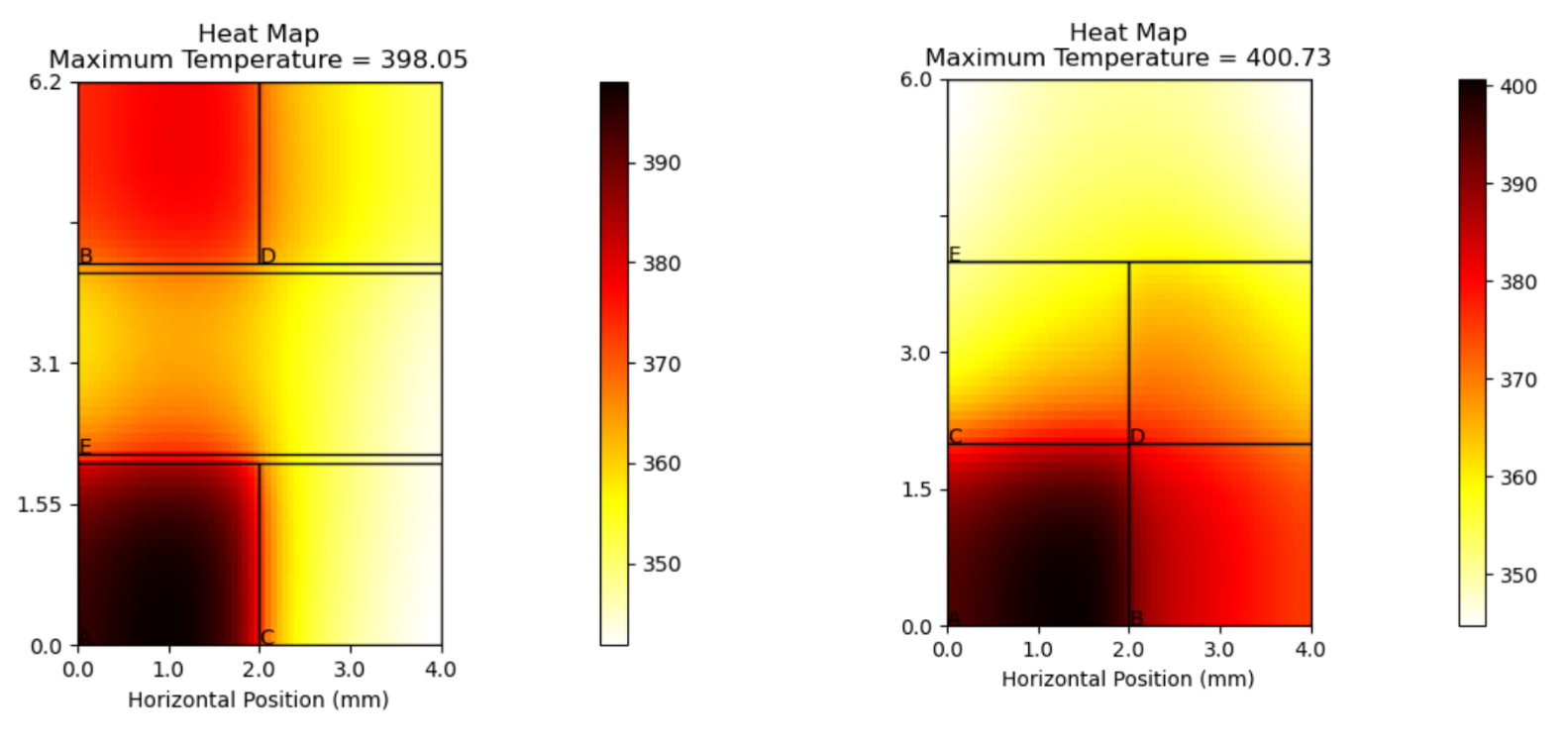}
    \caption{Validation of Thermal Efficiency of Chiplets over SoCs}
  \end{minipage}
\end{figure}

\subsection{Summary of Interconnect Packaging}

The text delves into semiconductor design and interconnect packaging technologies, focusing on chiplet integration, Eliyan's NuLink, the 2.5D RDL interposer, ASICLAND's advancements. 

\begin{table}[h]
\centering
\begin{tabular}{|p{3.5cm}|p{5cm}|}
    \hline
    \textbf{Technology/Innovation} & \textbf{Key Points} \\
    \hline
    Chiplet Integration & 
    \begin{itemize} 
        \item Heterogeneous integration with chiplets for efficient and flexible systems.
        \item Specific chiplet interface protocols cater to cores, analog components, and memory functionalities.
    \end{itemize} \\
    \hline
    Eliyan's NuLink & 
    \begin{itemize} 
        \item NuLink addresses limitations of silicon interposers.
        \item Mounts chiplets on a standard organic substrate for larger, cost-effective systems.
        \item Improves thermal performance and offers increased memory capacity.
    \end{itemize} \\
    \hline
    2.5D RDL Interposer & 
    \begin{itemize} 
        \item 2.5D RDL interposer offers cost-effective packaging for heterogeneous chip integration.
        \item Advantages include high productivity, lower costs, and reliability.
        \item Demonstrates robust performance in HPC applications.
    \end{itemize} \\
    \hline
    ASICLAND's Advancements & 
    \begin{itemize} 
        \item ASICLAND develops cost-efficient TSMC's CoWoS technology using RDL interposer.
        \item Incorporates a heat spreader for enhanced thermal management.
    \end{itemize} \\
    \hline
    Apple's M1 Ultra & 
    \begin{itemize} 
        \item M1 Ultra employs TSMC's CoWoS-S technology for chip connectivity.
        \item Uses ABF RDL and features a substrate supplied by Unimicron.
        \item Potential adoption of InFO LSI in the next M1 processor generation is speculated.
    \end{itemize} \\
    \hline
\end{tabular} \\
\caption{Summary: Interconnect Packaging Technologies}
\end{table}


\section{Thermal Analysis}
\subsection{Background}
The thermal management of chiplets brings forth numerous compelling advantages over System-on-Chip(SoCs). The most prominent is the former's modular design that allows for simplified thermal management with reduced impact on neighbouring elements. In contrast, SoCs necessitate the use of thermally inefficient floor planning simply due to the proximity of heat-generating components.
\subsubsection{Chiplet vs SoC}
Before moving on to thermally aware chiplet design, we must prove the thermal dominance of chiplets over traditional SoC architecture. This can be done through thermal simulation of two similar packages, one with SoC based design, the other subdivided into chiplets. 
\textbf{HotSpot v7.0}, a pre-RTL thermal simulator that supports the modelling of 2D, 2.5D and 3D integrated circuits (ICs) is used to demonstrate the same. Through LU Decomposition, it captures temperature traces of individual packages over time, relative to their power consumption.[39][42]

Simulation scenarios involved modelling a simple package over its native floor plan and then a more subdivided, chiplet-style floor plan under identical operating conditions. The steady-state temperatures over time reveal a notable decrease in maximum temperature by more than 2K in the chiplet-style configuration, affirming the enhanced thermal performance of chiplets.

\subsection{Design Considerations}
This subsection delves into the crucial aspects of the thermal management of chiplets, focusing both on production and packaging as well as external cooling methods. 

\subsubsection{Chiplet Packaging}
Owing to flexibility in the production state itself, chiplet placement can be optimized for numerous parameters such as peak temperature, area and cost. To substantiate the thermal prowess of chiplets, simulations were conducted using HotSpot v7.0, a thermally-aware micro-architecture simulation tool. 
The thermal efficiency of a packaging system plays a pivotal role in the overall performance and reliability of semiconductor devices. One key contributor to the same is the stack in which components are placed. Common stacks utilized today are 2D, 2.5D and 3D, each with their respective advantages and trade-offs:

\begin{itemize}
    \item \textbf{2D: }Traditional 2D packaging involves placing components on a single layer plane which proves to lower cost as well as complexity of production. It makes use of well-established manufacturing processes.

    \item \textbf{2.5D: }This packaging involves stacking of multiple dies or chiplets on a silicon interposer, to create a stacked structure to improve connectivity. The interposer allows for shorter interconnects, customizable floor planning as well increased surface area for thermal dissipation. 

    \item \textbf{3D: }This packaging involves stacking of multiple layers of active components directly on top of each other. The high interconnect density enables shorter signal paths, at the cost of concentrated thermal hotspots, and increased physical distance between cooling and heat-generating components.

\end{itemize}

\subsubsection{External Cooling}
Heat must be removed from the entire system outside the package, and transferred effectively and reliably to the ambient. While air cooling has proven to be a well-established low-cost option, it is inherently limited by surrounding conditions and low specific heat. Liquid cooling poses viable solutions for both these issues, but comes requires complex setups and necessitates adequate sealing and proofing. \textbf{Microfluidics} offers a promising solution in a blend of liquid and phase-change cooling that addresses heat transfer directly at its root while offering the most sought-after fluidic characteristics.

\subsection{Spec Information}
We start our pipeline by creating an initial floor plan based on a set of given specifications which will then be further optimized with thermal considerations and finally integrating the same with microfluidics and the corresponding cooling circuit. For a brief example, we have chosen the infotainment head unit for our demonstration herewith. To gain a basic specification and design baseline, we use the micro architecture diagram of the system and then collate the area and power requirement for each specific component in the architecture.

\begin{figure}[htbp]
  \centering
  \begin{minipage}[b]{0.5\textwidth}
    \centering
    \includegraphics[width=\textwidth]{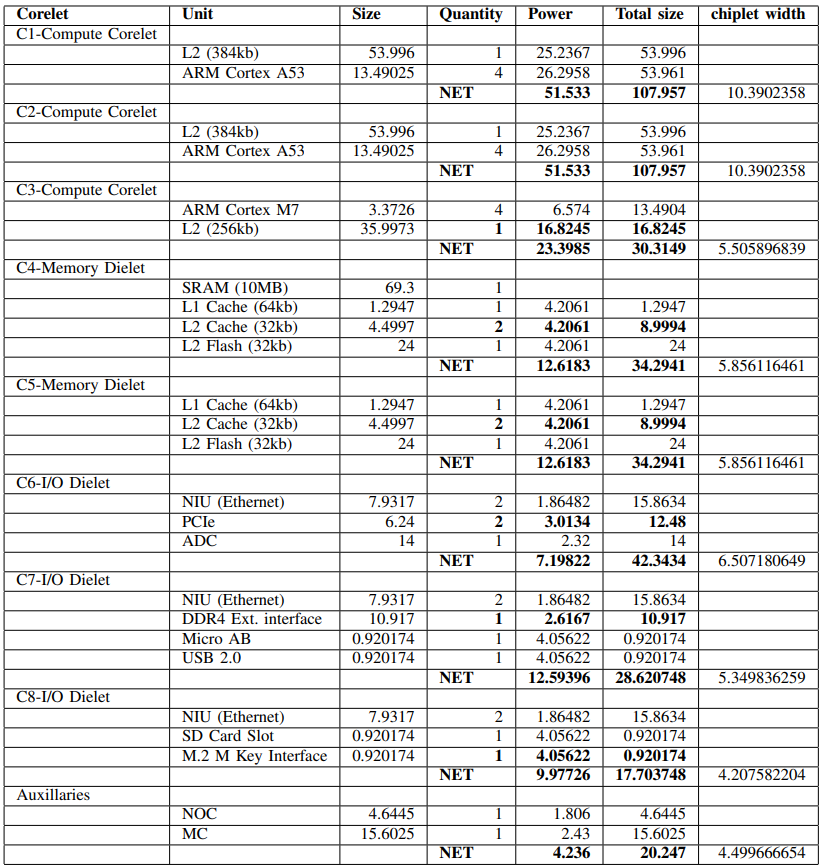}
    \caption{Spec for the Infotainment head architecture}
  \end{minipage}%
  \begin{minipage}[b]{0.5\textwidth}
    \centering
    \includegraphics[width=\textwidth]{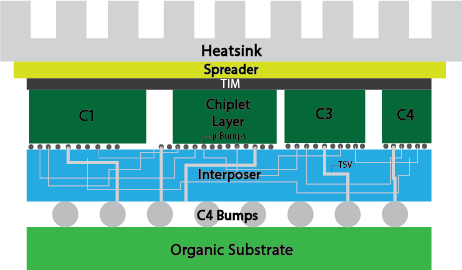}
    \caption{Cross-sectional view of typical 2.5D packaging architecture}
  \end{minipage}
\end{figure}

The design specifications were used to create the custom configuration file required to construct and optimize floorplans in our TAP 2.5D simulations. A floor plan with the respective material and thermal properties was used for the simulations.[40][41]

\subsection{Package Analysis}

A 2.5D package was chosen for the system with the following considerations:

\begin{itemize}
    \item \textbf{Enhanced Performance: }Shorter Interconnects through interposer lead to reduced latency and increased overall performance.
    \item \textbf{Thermals: }Larger surface area of interposer enhances heat dissipation
    \item \textbf{Cost-Effective} Passive interposer-based 2.5D integration with high yield has been proven commercially, drastically reducing system footprint.
\end{itemize}

A multi-layered sandwich approach incorporated the following

\begin{itemize}
  \item \textbf{Organic Substrate:} Provides structural support for the entire package and acts as a base for attaching various layers.

  \item \textbf{C4 Bumps (Controlled Collapse Chip Connection):} Facilitate electrical connections between the organic substrate and the interposer, enabling signal and power transmission between different layers.

  \item \textbf{Interposer with Through Silicon Vias (TSVs):} Serves as a bridge between chiplets and facilitates communication; it contains TSVs for vertical connections between different layers. A Passive Interposer is selected over active due to reduced thermal load, and simplicity of manufacturing as passive interposers use Back End of Line (BEoL) Manufacturing processes.

  \item \textbf{Microbumps:} Enable fine-pitch connections between the interposer and chiplet layer, ensuring high-density, high-bandwidth inter-chiplet communication.

  \item \textbf{Chiplet Layer:} Contains the individual chiplets (CPU, GPU, memory, etc.) and executes specific functions, contributing to the overall system functionality.

  \item \textbf{Thermal Interface Layer:} Manages and enhances thermal conductivity, ensuring efficient heat transfer from chiplets to subsequent layers.

  \item \textbf{Heat Spreader:} Disperses heat uniformly across the surface and enhances thermal performance by spreading heat generated by chiplets.

  \item \textbf{Heatsink:} Further dissipates heat into the surrounding environment, improving overall thermal management and preventing overheating. Provides an interface for interaction with fluid-cooling systems.
\end{itemize}

\subsubsection{Floorplan Optimization}

We leverage the customizability of floorplans in chiplet-style architecture to reposition and replace elements within the package, to optimize a \textit{cost function} with parameters like temperature, area, and cost. This is verified through extensive thermal simulation.
\begin{figure}[htbp]
     \includegraphics[width=0.6\textwidth]{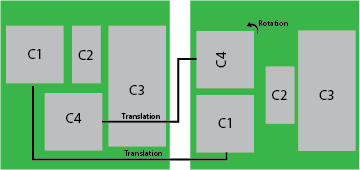} 
    \centering
     \caption{Translation and rotation of individual package elements} 
 \end{figure}

\begin{itemize}
    \item \textbf{Simulation Tools}
        \begin{itemize}
                           
            \item \textit{TAP-2.5D: } \textbf{TAP-2.5D} is an EDA tool that uses a simulated annealing-based algorithm to search for a chiplet placement solution that minimizes the total inter-chiplet wire length and the system temperature. It leverages HotSpot-6.0 for thermal simulation and a self-developed routing optimization tool (MILP) for wire length estimation.
            \begin{itemize}
                \item \textit{Cost Function- }
                A modified \textit{Cost Function} is employed to optimize inter-chiplet routing, minimizing both operating temperature and total wire length in a chosen \textbf{Infotainment Package}. It defines the simulated annealing cost function, normalizing temperature (T) and wire length (W) through Min-Max Scaling. The weights ($\alpha$  and 1-$\alpha$) represent the influence of temperature and length, dynamically adjusted by constraints to prioritize temperature reduction at higher temperatures ($\alpha > 0.5$) and wire length minimization below 85°C ($\alpha < 0.5$). This dynamic adjustment ensures a thermally feasible solution without compromising wire length optimization.
                \[
                    \text{Cost} = \alpha \times \frac{T - T_{\text{min}}}{T_{\text{max}} - T_{\text{min}}} + (1 - \alpha) \times \frac{W - W_{\text{min}}}{W_{\text{max}} -     W_{\text{min}}}
                \]

where

\[
\alpha =
\begin{cases} 
min{0.1 + \frac{T - 45}{100}, 0.9},  & \text{if } T > 60^\circ \text{C} \\
0, & \text{if } T \leq 60^\circ \text{C}
\end{cases}
\]

                \item \textit{Acceptance Probability- }
                The simulator makes use of a probabilistic algorithm that accepts subsequent chiplet placement based on a probability function given by 
                \[
\text{AP} = e^{\frac{\text{cost}(\text{current}) - \text{cost}(\text{neighbor})}{K}}
\]

A strategic decay K allows for reduced acceptance over successive iterations as the model approaches convergence and the cost of computation surpasses the benefit. 
            \end{itemize}
        \end{itemize}
\end{itemize}

\subsubsection{Simulation Results}

Using the above-mentioned tools, as well as the specification sheet in Fig. 18, a flp (Floorplan) was created for the \textbf{Infotainment Head Unit.}
\begin{itemize}
    \item \textbf{Annealing parameters: }
    Before setting up a floorplan optimization algorithm, it is important to setup basic annealing parameters for the same. This ensures that the simulation achieves a definite global optimum without extra computational cost. This was done by capping the number of iterations at a point where temperature tolerance exceeded subsequent improvement in the same.
    
    \textit{Convergence Criterion}
    
    \[|T_i - T_{i-1}| < \text{Tol}\]
\\
    Numerous iterations were run with step changes to the K value and points were curve-fitted for better visualization. 

\begin{figure}[htbp]
  \centering
  \begin{minipage}[b]{0.5\textwidth}
    \centering
    \includegraphics[width=\textwidth]{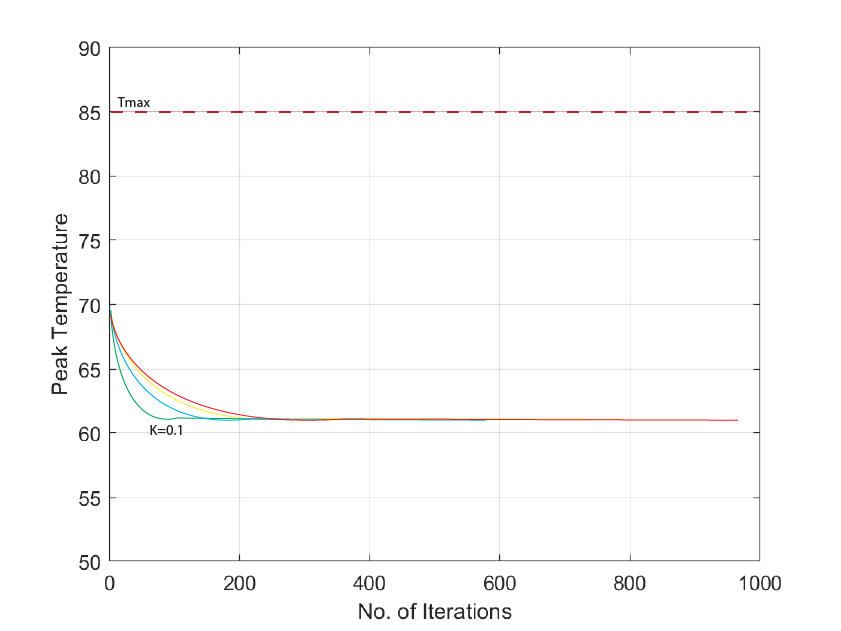}
    \caption{Determination of Ideal K using Temperature Tolerance Criteria}
  \end{minipage}%
  \begin{minipage}[b]{0.5\textwidth}
    \centering
    \includegraphics[width=\textwidth]{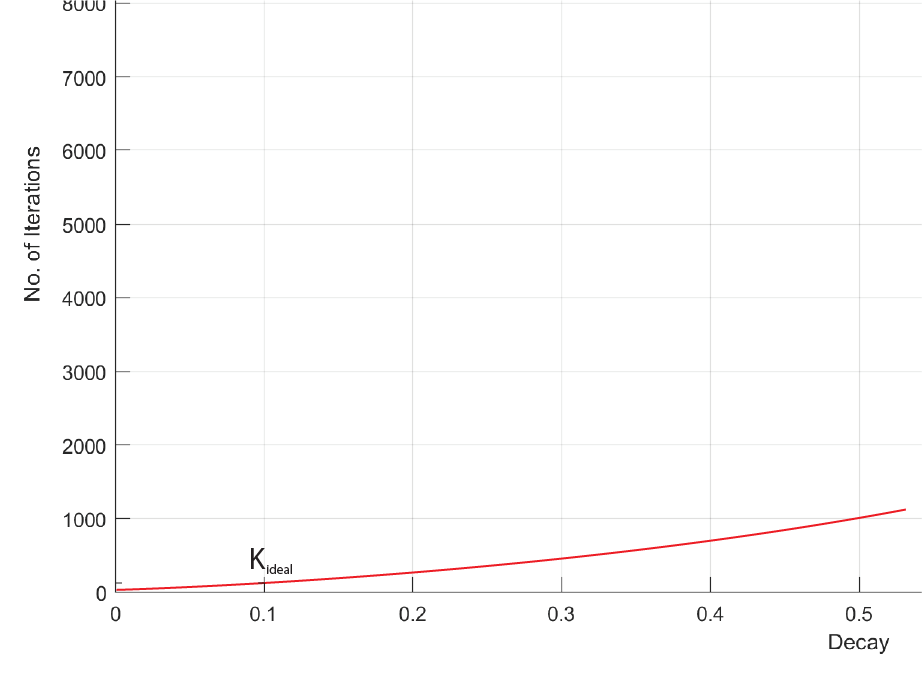}
    \caption{Number of iterations for Ideal K value}
  \end{minipage}
\end{figure}

   The hence established K-values were then fed into the \textit{Acceptance Probability} criterion to generate an educated thermal model.

    Coming to the \textit{Cost Function}, we iterated through stepped-down interposer areas to model the impact of available area on peak temperatures on the chiplet layer. 
       
\begin{figure}[htbp]
  \centering
  \begin{minipage}[b]{0.5\textwidth}
    \centering
    \includegraphics[width=\textwidth]{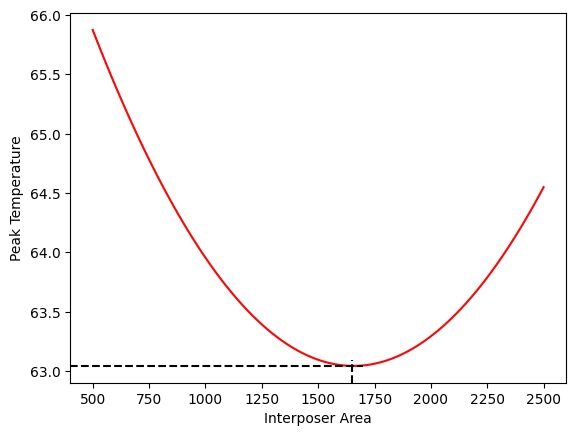}
    \caption{Impact of Interposer Area on Peak Chiplet Temperatures}
  \end{minipage}%
  \begin{minipage}[b]{0.5\textwidth}
    \centering
       \includegraphics[width=\textwidth]{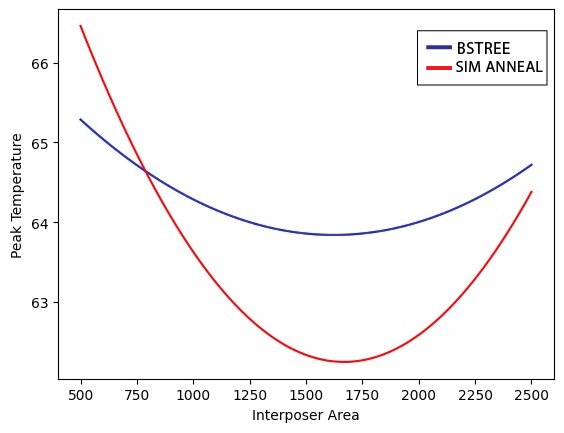}
     \caption{Number of iterations for Ideal K value} 
  \end{minipage}
\end{figure}

 Datapoints were once again collected for discreet viable interposer dimensions after which they were fed into a polyfit function. A \textbf{minima} for peak temperature was derived at an area of \textit{1600$cm^2$} or an interposer of dimensions 40x40mm.

 \item \textbf{Implementation: }
 Peak Temperature trends were noted for an initial \textit{Binary Search Tree} Placement and compared to an optimized floorplan through derived through the \textit{Custom Simulated Annealing} algorithm. A stark drop in peak temperatures post optimization backs the strategic placement of components on the interposer.

\begin{figure}[htbp]
  \centering
  \includegraphics[width=0.6\textwidth]{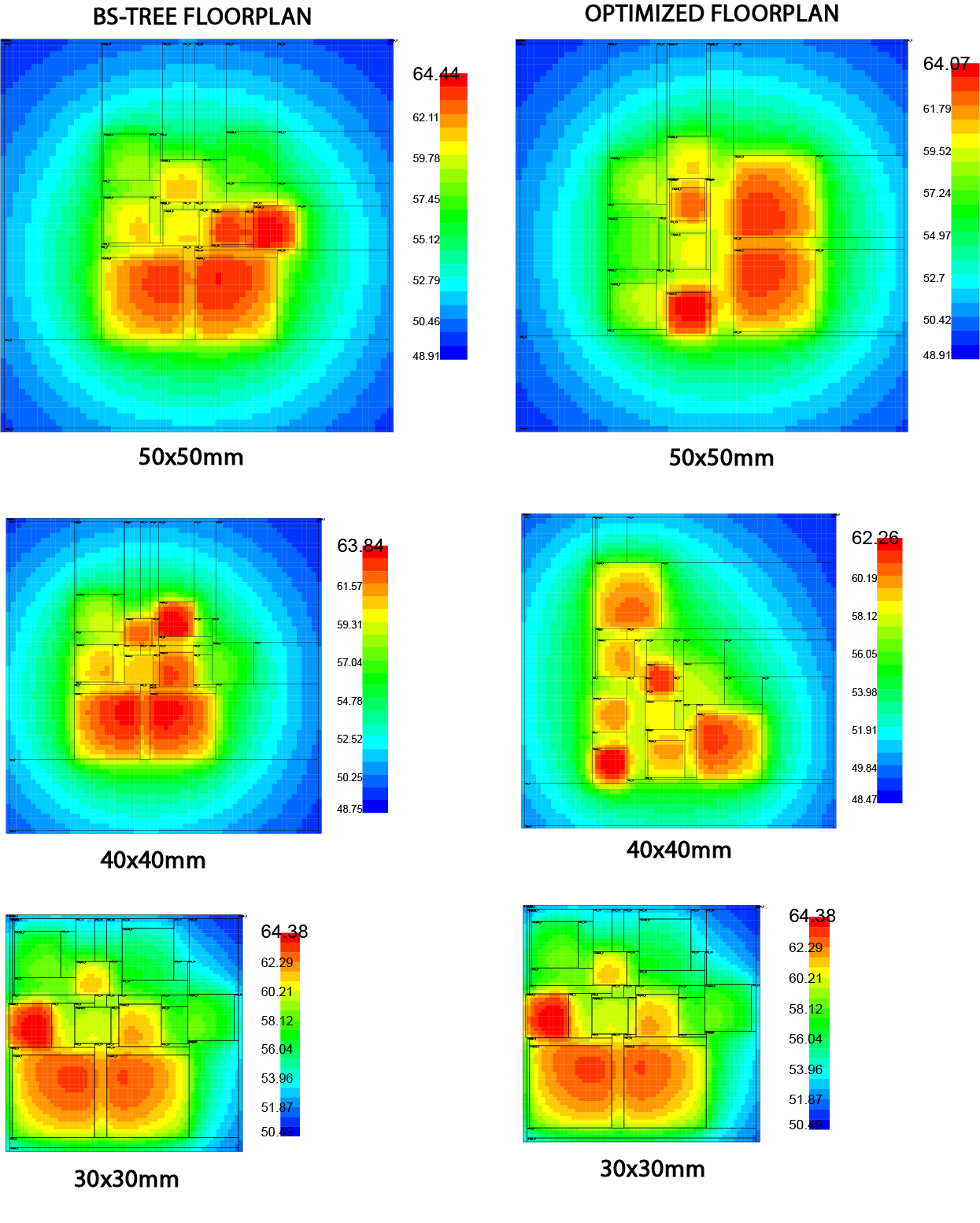}
  \caption{Comparison of Peak temperatures - Pre and Post Optimized Placement}
\end{figure}
 
     Based on the above-drawn conclusions, the floorplan, along with appropriate power traces of individual components was fed into TAP-2.5D. After the predicted ~110 iterations, a completed \textit{optimized floorplan} was formed and outputted as an SVG file.
    
\end{itemize}
\subsection{Cooling Approach}
The increased integration density of electronic components and subsystems, including the nascent commercialization of 3D chip stack technology, has exacerbated the thermal management
challenges facing electronic system developers. The confluence of chip power dissipation above 100 W, localized hot spots with fluxes above 1 kW/cm2, and package-level volumetric heat
generation that can exceed 1 kW/cm3 has exposed the limitations of the current “remote cooling” paradigm and its inability to support continued enhancements in the performance of advanced silicon and compound semiconductor components. To overcome these limitations and remove a significant barrier to continued Moore’s law progression in electronic components and systems, it is essential to “embed” aggressive thermal management in the chip, substrate, and/or package and directly cool the heat generation sites.\cite{b10}

One of our main objectives in analyzing our cooling methodology was to ensure the safe operation of all systems even during the case of a failure in the cooling system. This translated to having a safe threshold temperature when simulating the floorplans for optimal placement and also ensuring the passive cooling through the immersion cooling system further guaranteed the safety factor of that threshold. In addition to the safety aspect, the system is designed around the sole purpose of deploying it in automotive systems and use cases and needs to have the robustness to handle the conditions subjected to in its life cycle and ensure optimal performance, often in scenarios and boundary conditions where fundamental assumptions might fail ( example: assumption of available draft air or forced convection for passive heatsinks or cooling loops with radiators. ). An additional facet of the approach was to ensure maximum effectiveness of the system while minimizing power requirements for the same as building the system centered around an EV should be kept in mind and the energy cost must be present in all considerations. These conditions led us to choose microfluidics with a two-phase passive cooling loop as our best option for the cooling architecture. We will delve into the same below.  
\subsection{Microfluidics}

\begin{figure}[htbp]
  \centering
  \begin{minipage}[b]{0.5\textwidth}
    \centering
    \includegraphics[width=\textwidth]{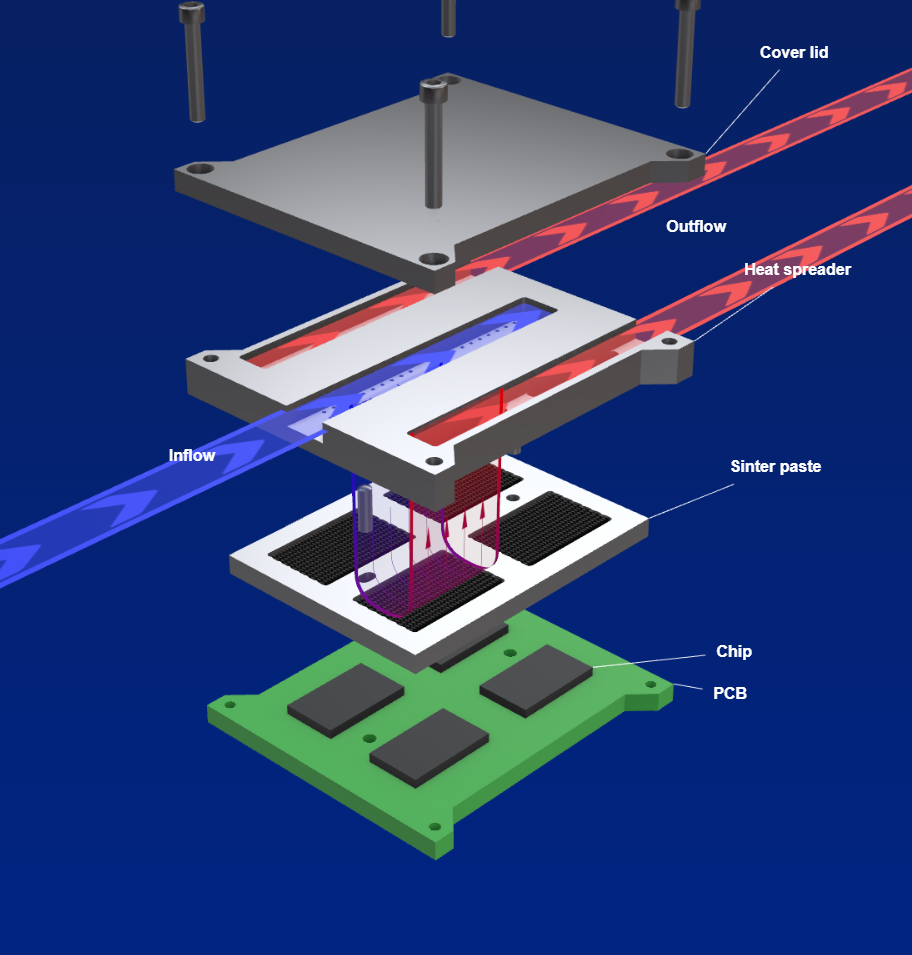}
    \caption{Exploded view of the microfluidics design \cite{b12}    }
  \end{minipage}%
  \begin{minipage}[b]{0.5\textwidth}
    \centering
    \includegraphics[width=\textwidth]{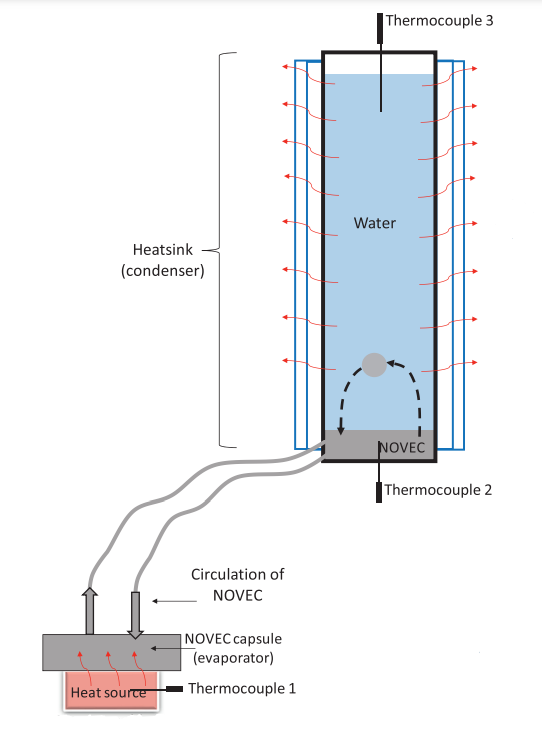}
    \caption{ Two-phase passive cooling system}
  \end{minipage}
\end{figure}

Microfluidics was chosen as the best choice for the cooling architecture as it is the most efficient method to remove heat directly from the source of heat at the interposer level in the chiplet architecture. Power electronics are solid-state electronic devices that convert electrical power into different forms, and are used in a vast array of daily applications ( from computers to battery chargers, air conditioners to hybrid electric vehicles, and even satellites ). The rising demand for increasingly efficient and smaller power electronics means that the amount of power converted per unit volume of these devices has increased dramatically. This, in turn, has increased the heat flux of the devices ( the amount of heat produced per unit area ). The heat generated in this way is becoming a big problem. Miniaturized electronic devices generate a lot of heat, which must be dissipated to maintain performance. \cite{b13} 

A microfluidics system designed to be an integral part of a chiplet ecosystem demonstrates exceptional cooling performance. This not only provides us with the flexibility in designing the most effective microfluidics cooling loop centered around the thermal hotspots in our chiplet, but the passive cooling loop with two-phase cooling ensures the maximum possible steady state temperature of all regions in the chiplet is always within the safe threshold. Other alternatives that use forced convection ( either liquid or air ) are both less efficient in extracting heat from the source and are required to have accommodations for airflow or a radiator. Our microfluidic channels use a two-phase cooling loop ( consisting of Novec 7000 and water )     
\subsection{Two-Phase Passive Immersion Cooling}

 A two-phase passive immersion cooling loop was chosen over a single-phase system as it enables us to reach steady-state operating conditions of the chiplet faster for extreme and oscillating loads through the rapid vaporization and condensation cycle of Novec. Compared to single-phase cooling, two-phase cooling provides improved heat dissipation for a given volume or mass of fluid because the latent heat of a fluid can be orders of magnitude larger than the specific heat. This immersion cooling setup helps make the system more efficient, reduces energy consumption, and minimizes costs when compared to a traditional liquid cooling system.

 Traditionally, electronics have relied on air-cooled heat sinks or liquid-cooled cold plates to
manage electronic waste heat. However, cooling schemes using liquid-vapor phase change
(hereafter referred to as “two-phase cooling”) are the practical and cost-conscious step beyond single-phase cooling. The US Department of Defense (DoD) and the National Aeronautics and Space Administration (NASA) have distinguished two-phase cooling as a favourable solution to meeting the strict demands for some emerging cooling in their automotive platform requirements.\cite{b14}
 
 The microfluidic channels are connected to a condensing container with water where the evaporated Novec instantly cools down into its liquid form by giving out heat as the bubbles pass through the water column. The heat accumulated in the water is further lost to the environment passively through the container which itself acts as a heatsink\cite{b11}. The rapid transportation of heat through Novec vapor bubbles ensures there is effective heat flux even though the entire system works passively without the need for an external head or pump to drive the loop. The system uses the low evaporation temperature of NOVEC to increase the cooling efficiency and buoyancy flow of vapour to produce quick fluid movement. The condenser is based on a water pool in which the NOVEC bubbles are efficiently condensed without the effect of non-condensable gases and the water suffers an important agitation that increases the heat transfer to
the walls.
 
\section{Conclusion}

The Chiplets in totality outperform the traditional monolithic SoC design be it compute-level optimization, task parallelisation, power consumption and maximising the golden ratio of compute. The Gem5 simulation results for micro-architecture design, thermal Tap-2.5D simulation results for SoC vs chiplets and inter chiplet placement optimization , the suggestive Forced convection cooling technique also validate the very fact that Chiplets are the future of automotive range hardware .A comprehensive comparison of UCIe, AIB, BoW, QPI, Infinity Fabric, and LIPINCON and physical layer parameter analysization recommend UCIe for digital interfaces, HBM2 and HBM3 for digital-memory communication, and BoW for analog-digital interfaces. Security and reliability considerations align with ISO 21434 standards, incorporating features like secure boot, hardware security modules, and continuous monitoring for chiplet-level security. Exploration of NuLink and RDL interposer as alternative interconnect packaging technologies provides flexibility aligned with specific requirements.Therefore, chiplets represent the optimal solution to address the high computational demands in today's rapidly evolving environment.

\bibliographystyle{unsrt}  
\bibliography{references}

\end{document}